\documentclass[10pt,journal,compsoc]{IEEEtran}
\ifCLASSOPTIONcompsoc
  \usepackage[nocompress]{cite}
\else
  \usepackage{cite}
\fi
\ifCLASSINFOpdf
\else
\fi
\usepackage{amsmath,amssymb,amsfonts}
\usepackage{graphicx}
\usepackage{textcomp}
\usepackage{xcolor}
\usepackage{comment}

\usepackage[linesnumbered,vlined,boxed,commentsnumbered]{algorithm2e}
\usepackage{setspace}
\usepackage[parfill]{parskip}

\begin{document}

\title{Bug-locating Method based on Statistical Testing for Quantum Programs}

\author{Naoto~Sato and Ryota~Katsube % <-this % stops a space
\IEEEcompsocitemizethanks{\IEEEcompsocthanksitem N. Sato and R. Katsube are with Research \& Development Group, Hitachi, Ltd., Japan.\protect\\
% note need leading \protect in front of \\ to get a newline within \thanks as
% \\ is fragile and will error, could use \hfil\break instead.
E-mail: naoto.sato.je@hitachi.com
%\IEEEcompsocthanksitem J. Doe and J. Doe are with Anonymous University.
}% <-this % stops an unwanted space
%\thanks{Manuscript received April 19, 2005; revised August 26, 2015.}
}

\IEEEtitleabstractindextext{%
\begin{abstract}
When a bug is detected by testing a quantum program on a quantum computer, we want to determine its location to fix it. To locate the bug, the quantum program is divided into several segments, and each segment is tested. However, to prepare a quantum state that is input to a segment, it is necessary to execute all the segments ahead of that segment in a quantum computer. This means that the cost of testing each segment depends on its location. We can also locate a buggy segment only if it is confirmed that there are no bugs in all segments ahead of that buggy segment. Since a quantum program is tested statistically on the basis of measurement results, there is a tradeoff between testing accuracy and cost. These characteristics are unique to quantum programs and complicate locating bugs. 
%We suggest that these characteristics should be considered to efficiently locate bugs. 
We propose an efficient bug-locating method consisting of four approaches, cost-based binary search, early determination, finalization, and looking back, which take these characteristics into account. We present experimental results that indicate that the proposed method can reduce the bug-locating cost, represented as the number of executed quantum gates, compared to naive methods that do not use the four approaches. The limitation and usefulness of the proposed method are also discussed from the experimental results.
\end{abstract}

% Note that keywords are not normally used for peerreview papers.
%\begin{IEEEkeywords}
%testing and debugging, quantum software, probability and statistics.
%\end{IEEEkeywords}
}

% make the title area
\maketitle

% To allow for easy dual compilation without having to reenter the
% abstract/keywords data, the \IEEEtitleabstractindextext text will
% not be used in maketitle, but will appear (i.e., to be "transported")
% here as \IEEEdisplaynontitleabstractindextext when the compsoc 
% or transmag modes are not selected <OR> if conference mode is selected 
% - because all conference papers position the abstract like regular
% papers do.
\IEEEdisplaynontitleabstractindextext
% \IEEEdisplaynontitleabstractindextext has no effect when using
% compsoc or transmag under a non-conference mode.

% For peer review papers, you can put extra information on the cover
% page as needed:
% \ifCLASSOPTIONpeerreview
% \begin{center} \bfseries EDICS Category: 3-BBND \end{center}
% \fi
%
% For peerreview papers, this IEEEtran command inserts a page break and
% creates the second title. It will be ignored for other modes.
\IEEEpeerreviewmaketitle

%-----------------------------------------------------------------------
\IEEEraisesectionheading{\section{Introduction}\label{intro}}
The field of quantum software engineering has developed rapidly \cite{piattini2020talavera} \cite{piattini2021quantum} \cite{survey5} \cite{moguel2020roadmap} \cite{zhao2020quantum} \cite{DESTEFANO2022111326} \cite{garcia2023quantum}, and research on testing, verification, and debugging of quantum programs began with a typology of bugs \cite{huang2018qdb} \cite{bugpattern}. On the basis of these studies, an application of classical methods to quantum programs was proposed \cite{classicalTest}. 
When a bug is detected by testing a quantum program on a quantum computer, we divide the program into several segments and test each segment to determine the detailed bug location. To test each segment, it is necessary to prepare the quantum states that would be input to each segment when the entire quantum program is executed. To prepare the input quantum state for a segment in a quantum computer, all segments ahead of that segment should be executed on the initial state that the quantum computer physically forms \cite{Long2022lwc}.
This leads to the cost of testing segments to depend on the locations of the segments. Even if a bug is detected in the test of a segment, it does not necessarily mean that the bug is in that segment. This is because the bug may be in other segments ahead of that segment. 
Therefore, to locate a buggy segment on a quantum computer, we have to confirm that there is no bug in any segment ahead of it. Another perspective is that the testing of each segment is conducted on the basis of measurements. Since a sufficient number of measurements is necessary for testing with sufficient accuracy, there is a tradeoff between testing accuracy and its cost.

In our prior study \cite{sato2024locating}, we introduced these characteristics unique to quantum programs as factors that make it difficult to locate a buggy segment. We also introduced four approaches to efficiently locate a buggy segment on the basis of these characteristics. In this paper, with reviewing those characteristics and approaches. we update one of the approaches and present a deteiled method consisting of those approaches. We also demonstrate the efficiency of the proposed method more rigorously with experimental results.
%; however, they have not been mentioned in previous studies. 
%The first contribution of this paper is clarifying for the first time the characteristics that should be considered to efficiently locate a buggy segment in quantum programs on a quantum computer. The second contribution is that we propose a bug-locating method for quantum programs. We implement the proposed method and conducted experiments to demonstrate its efficiency.

In Section \ref{background}, we briefly introduce quantum programs and their testing. Related work on quantum program testing and debugging is then described in Section \ref{background}. We present the characteristics that should be considered to efficiently locate a buggy segment in quantum programs on a quantum computer in Section \ref{chara}. On the basis of those characteristics, we present the proposed bug-locating method which consists of the above four approaches in Section \ref{proposed}. We demonstrate the efficiency of the proposed method from experimental results in Section \ref{exp}. In Section \ref{discuss}, we discuss the limitations and usefulness of the proposed method. We draw conclusions from this study and present future work in Section \ref{conclusion}.

\section{Background}\label{background}
\subsection{Quantum Program}
A qubit, which is a variable in a quantum program, can be in a superposition of the basis states |0> and |1>. When a qubit is measured, 0 or 1 is observed probabilistically, depending on its state. The state of a qubit can be expressed as $|\psi> = a_0 |0> + a_1 |1>$ ($|a_0|^2 + |a_1|^2=1$), where $a_0$ and $a_1$ are complex numbers and called amplitudes. The absolute squares of the amplitudes $|a_0|^2$ and $|a_1|^2$ represent the probabilities of obtaining 0 and 1 from a Z-basis measurement. If a quantum state is measured many times, the absolute square of the amplitude can be estimated from the measurement results. An arbitrary quantum state consisting of $n$ qubits is generally represented by $2^n$ basis states. For example, a two-qubit state consists of a superposition of $|00>$, $|01>$, $|10>$, and $|11>$.

\begin{figure}[!t]
\centering
\scalebox{0.60}{\includegraphics[bb=0 0 217 81]{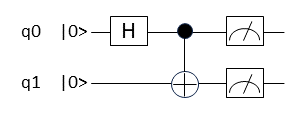}}
\caption{Quantum circuit to create Bell state}
\label{bell}
\end{figure}

A quantum program is represented using a model called a quantum circuit. Figure \ref{bell} shows an example of a quantum circuit to generate the Bell state and measure it. Each horizontal line corresponds to a qubit, and the operations on them, called quantum gates, are arranged from left to right. 
In this quantum circuit, qubits $q0$ and $q1$ are both initialized to $|0>$. After the initialization, the Hadamard gate is executed for $q0$, then the CNOT gate is executed on $q0$ and $q1$. The quantum state formed after these executions is $|\psi> = \frac{1}{\sqrt{2}} |00> + \frac{1}{\sqrt{2}} |11>$, which is called the Bell state. When $q0$ and $q1$ in the Bell state are measured, 00 or 11 is observed. Since the amplitude of $|00>$ is $ \frac{1}{\sqrt{2}}$, the probability of obtaining 00 is $ |\frac{1}{\sqrt{2}}|^2=\frac{1}{2}$. Similarly, the probability of obtaining 11 is $ \frac{1}{2}$. %The absolute squares of the amplitudes of $|00>$ and $|11>$ can be estimated from the measurement results.

\subsection{Quantum Program Testing}\label{test}
%量子プログラムのバグを検出する方法としては、古典プログラムと同様に、テスティングが有効である。量子プログラムのテストには、古典コンピュータ上でシミュレータを使用する場合もある。しかし、古典シミュレーションでは、$n$量子ビットからなる量子プログラムを、$2^n \times 2^n$の行列計算プログラムとして扱うため、量子ビットの数が多い大規模な量子プログラムの場合は、計算で扱う行列のサイズが膨大となり、実行困難になる。その場合、量子コンピュータを用いてテストを行うことになる。
Testing is an effective method for detecting bugs in quantum programs as well as in classical programs. For testing quantum programs, a simulator running on a classical computer is useful. Since an $n$-qubit quantum program is treated as a matrix calculation program of $2^n \times 2^n$, it is difficult to test a quantum program with a large number of qubits in a classical simulation. In such cases, a quantum computer is necessary for testing.

We assume a quantum program so large that it cannot be simulated in practical time on a classical computer. Since the motivation for using a quantum computer is to solve complex problems that cannot be solved with a classical computer, this assumption is natural.

Long et al. classified testing methods for quantum programs into statistic-based detection (SBD) and quantum runtime assertion (QRA) \cite{Long2022lwc}. When executing a quantum program on a quantum computer, it is not possible to directly read the quantum state of qubits \cite{DiMatteo:2024thh}. With SBD methods, the output quantum state is measured many times. Quantum information derived from the measurement results is then statistically compared with the test oracle. If they are different, the quantum program is statistically determined to have a bug. A simple SBD method is directly comparing the measurement results with the expected results that can be derived from the absolute squares of the expected amplitudes.
%evaluating the absolute squares of the amplitudes that can be directly derived from the measurement results \cite{huang2019statistical}. 
For a more rigorous test, the density matrix may be useful, which is calculated by quantum state tomography, maximum likelihood estimation, or Bayesian estimation \cite{nielsen2010quantum}\cite{TANAKA20122471}\cite{lukens2020practical}.

Since QRA methods were proposed for runtime testing, they do not destroy the tested quantum state and can be conducted using the result of a single measurement \cite{li2020projection}\cite{liu2020quantum}\cite{runtime2021}. 
However, QRA methods require the embedding of an extra complex program in the tested program. The implementation of the extra program is not easy, and if there is a bug in the extra program, the test cannot be conducted correctly \cite{Long2022lwc}. Therefore, if we do not want to run the tests at runtime, SBD methods are still useful.

\subsection{Related Work} \label{relwork}
Various SBD methods have been proposed by applying classical testing methods to quantum programs. Huang et al. proposed a statistical assertion testing method based on the chi-square test \cite{huang2019statistical}. Honarvar et al. proposed a property-based testing method with which the pre- and post-conditions of a quantum program are described by assertions \cite{honarvar2020property}. Ali et al. proposed the input state space of the quantum program as input coverage. Similarly, variations in the measurement results is suggested as output coverage. They also proposed a tool that generates test cases on the basis of those coverage metrics \cite{assessing}\cite{wang2021quito}. An approach based on metamorphic testing, which has recently been applied to the testing of artificial intelligence software \cite{chen2018metamorphic}, has also been proposed for application to quantum programs \cite{abreu2022metamorphic} \cite{MorphQ}. Long and Zhao defined unit and integration tests in quantum programs and proposed a method of equivalence class testing for quantum programs \cite{Long2022lwc}. 
%Muqeet et al. proposed leveraging machine learning to improve the accuracy of SBD testing methods under the noise effect \cite{muqeet2023noise}. 
Other classical software testing methods including fuzzing \cite{wang2021poster}, mutation testing \cite{mendiluze2021muskit}\cite{fortunato2022qmutpy}, search-based testing \cite{wang2022qusbt}\cite{wang2022mutation}, and combinatorial testing \cite{wang2021application} have been proposed for application to quantum programs. Our proposed method uses an SBD method internally.%, but it does not depend on the details of the testing method. In our experiments discussed in Section \ref{exp}, we use the chi-square test as a basic testing method \cite{huang2019statistical} which compares the measurement results with the expected results.

Li et al. presented runtime assertion as a QRA method involving a projective measurement that stabilizes the tested quantum state \cite{li2020projection}. Liu et al. introduced another runtime assertion by adding extra (ancilla) qubits to collect the information of the tested quantum state \cite{liu2020quantum}\cite{runtime2021}.

With regard to debugging quantum programs, Miranskyy et al. discussed the applicability of the classical debugging strategies to quantum programs \cite{miranskyy2020your} \cite{miranskyy2021testing}. Liu et al. suggested using the information obtained from assertion tests for debugging \cite{liu2020quantum}.  Zhao et al. introduced bug patterns of quantum programs and a static analysis tool based on the patterns \cite{zhao2023qse}. Li et al. proposed a debugging scheme to locate bugs by injecting assertions into a quantum program \cite{li2020projection}. They also suggested that to show that a segment has a bug, it is necessary to confirm that all segments ahead of it do not have a bug. However, it was not recognized as a factor that complicates bug locating, and their search strategy was based on naive linear search. This paper clarifies four characteristics that should be considered to locate bugs and presents our efficient bug-locating method that takes these characteristics into account.

Metwalli et al. suggested that circuit slicing, in which the quantum program is divided into smaller segments, is efficient for testing and debugging \cite{metwalli2023cirquo}. They also defined the types of segments on the basis of their algorithmic functions and presented different testing and debugging strategies for each type. However, they do not refer to how to locate a bug included in any of the segments. 
Visual assistance is also important for debugging. One begins with the confirmation of the overview of the program then moves to a detailed review of the sub-components. Quantivine, a visualization tool for large quantum programs, was developed to visually assist the drill down by expanding the sub-components \cite{wen2023quantivine}. Toward fixing detected bugs, Guo et al. suggested that ChatGPT is useful to automatically repair quantum programs \cite{guo2024repairing}.

\section{Characteristics of Quantum Program Testing for Locating Bugs}\label{chara}
\begin{figure*}[hbt]
\centering
\scalebox{0.25}{\includegraphics[bb=0 0 2063 822]{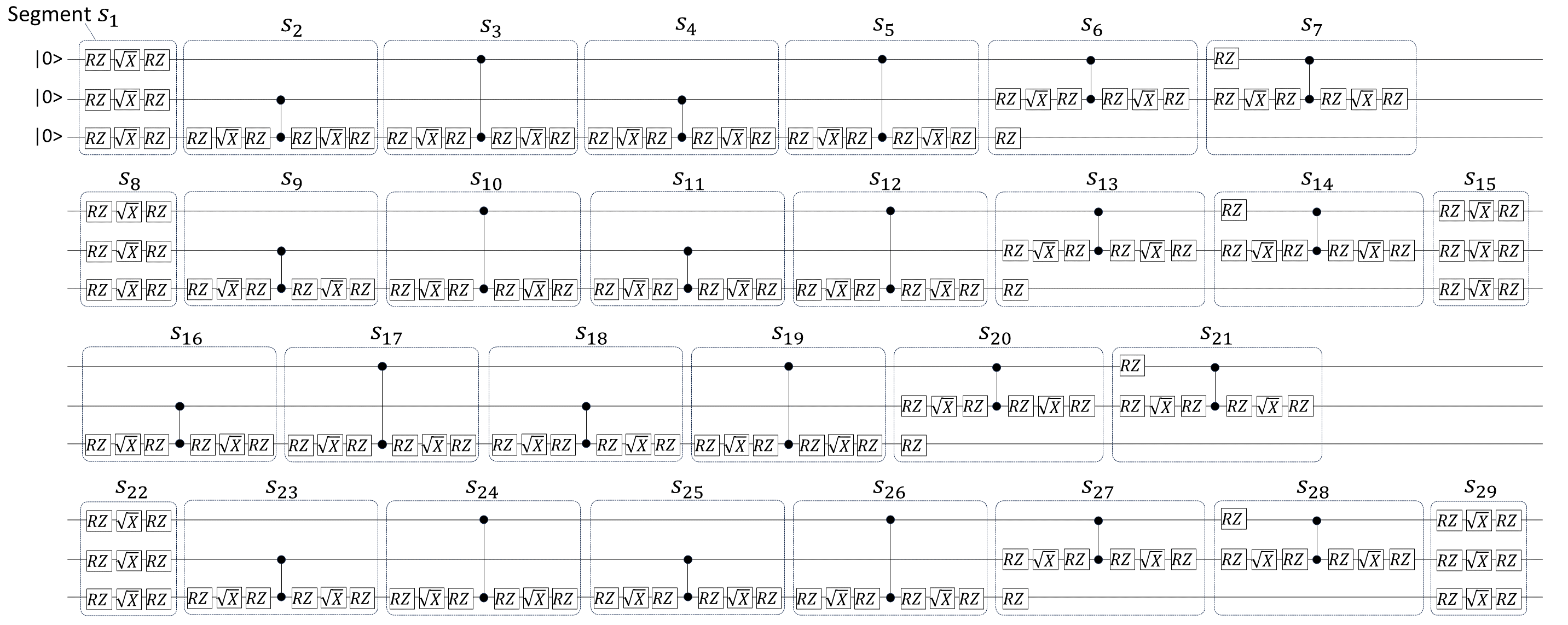}}
\caption{Example of quantum program divided into segments}
\label{segmentation}
\end{figure*}

When a bug is detected in a quantum program, the program is divided into several segments \cite{metwalli2023cirquo}, and each segment is tested to determine the bug location. For instance, we use a simple example program implementing Grover's algorithm \cite{groverexample}, where the solution being searched for is 111. This program can be divided into 29 segments, as shown in Figure \ref{segmentation}. This consists of $CZ$, $RZ$, and $ \sqrt{X}$ gates, which are the native gates of IBM Quantum\texttrademark Heron. The rotation angles of $RZ$ gates are omitted in this figure.

\begin{figure}[hbt]
\centering
\scalebox{0.25}{\includegraphics[bb=0 0 754 430]{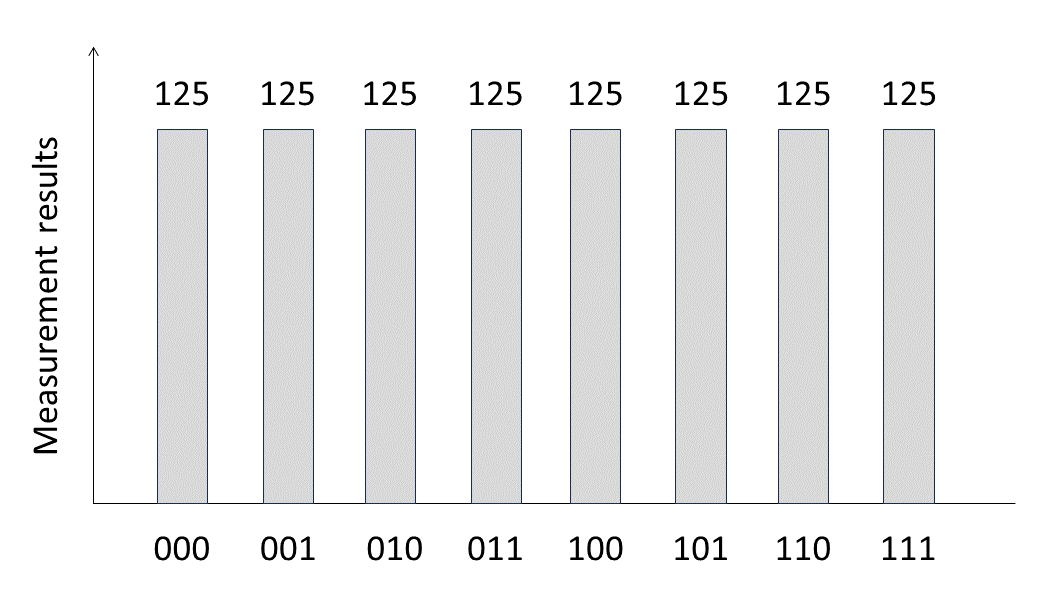}}
\caption{Categorical distribution representing expected output state of $s_1$}
\label{distribution}
\end{figure}

In testing each segment, we assume that the developer can expect the correct output quantum state of each segment as a test oracle. In the example in Figure \ref{segmentation}, the output state of segment $s1$ is expected to be a uniform superposition state. In our study, we used the chi-square test for testing each segment, as in a previous study \cite{huang2019statistical}. The chi-square test statistically evaluates the results of Z-basis measurements compared with its test oracle. The oracle of the measurement results is calculated from the expected amplitudes and number of measurements. 
%absolute squares of the amplitudes estimated from the results of Z-basis measurements with its oracle. 
For example, the test oracle of $s_1$ is the expected measurement results of the output state. It is represented by a categorical distribution, as shown in Figure \ref{distribution}.
This distribution is derived from the expected values of the absolute squares of the amplitudes, which are 0.125 for all bases. If 1000 measurements are conducted, the distribution in Figure \ref{distribution} is calculated as a test oracle.
Similarly, since only the phases of some bases are inverted in $s_2$, the same distribution as in Figure \ref{distribution} is expected as the oracle of $s_2$. %In the following, we show four characteristics of quantum program testing that make it difficult to locate a buggy segment.

To test each segment, the quantum states that would be input to each segment when the entire quantum program is executed should be used as test input. However, unlike a classical computer, a quantum computer cannot easily prepare the quantum state as desired. A quantum computer always initializes a quantum state to a unique device-specific state, which is often $|0>^{\otimes n}$, where $n$ is the number of qubits. Therefore, if we want to prepare the input state of a tested segment, we need to execute several quantum gates from the initial state $|0>^{\otimes n}$. These quantum gates executed to prepare the input state correspond to the gates ahead of the segment in the entire quantum program.
This means that when testing a backward segment, more quantum gates are executed than when testing a forward segment. In the testof segment $s_{20}$ in Figure \ref{segmentation}, segments $s_1$ to $s_{19}$ are executed to prepare the input state of $s_{20}$. In the test of segment $s_2$, however, only segment $s_1$ is executed to prepare the input state of $s_2$. The first characteristic \textbf{[C1]} is that the cost of testing a segment depends on its location.

It is also implied that if we test a segment and a bug is detected, it is possible that the bug is not in the segment but in another segment ahead of it. For example, if a bug is detected in the test of segment $s_{20}$ in Figure \ref{segmentation}, this indicates that there is a bug from segment $s_1$ to $s_{20}$. If we want to locate the bug in $s_{20}$, it is necessary to confirm that there is no bug from $s_1$ to $s_{19}$. Therefore, the second characteristic \textbf{[C2]} is that we can locate a buggy segment only if it is confirmed that there is no bug in all segments ahead of the buggy segment.

This leads to another characteristic. If a bug is not detected in the test of a segment, it can be assumed that the segment ahead of it is also bug-free (more precisely, no bug affecting the output of the segment). Conversely, if a bug is detected, the same bug should also be detected in the tests of the segments behind it. In the same example, assume that a bug is not detected in the test of $s_{20}$. It also shows that $s_1$ to $s_{19}$ do not have a bug (affecting the output state of $s_{20}$). If a bug is detected in the test of $s_{20}$, that bug will be detected in the tests of $s_{21}$ to $s_{29}$ (unless there is another bug that cancels out the effect of the first bug). Accordingly, the third characteristic \textbf{[C3]} is that the test result of a segment may be predicted from those of other segments.
%the test results of the segments are not independent.

As described in Section \ref{test}, the testing of each segment is statistically conducted on the basis of the measurement results. A sufficient number of measurements is necessary for testing with sufficient accuracy. This indicates the fourth characteristic \textbf{[C4]} that there is a tradeoff between testing accuracy and its cost. For example, for arbitrary quantum state $ |\psi> = a_0 |0> + a_1 |1> $, the absolute square of the amplitude $a_0$, that is, ${| a_0 |}^{2}$, is estimated from the measurement results. The standard deviation $ \sigma $ of the maximum likelihood estimator of ${| a_0 |}^{2}$ is expressed as $ \sigma = \sqrt{\frac{{| a_0 |}^{2} (1 - {| a_0 |}^{2})}{M}}$ \cite{rebentrost2018quantum}, where $M$ is the number of measurements. This means that the accuracy of the estimation depends on $M$.

In the chi-square test that we use in this study, accuracy is indicated by the p-value and power of the test. The power described as $1 - \beta $ is the probability of correctly detecting the presence of a bug, where $ \beta $ is the Type I\hspace{-1.2pt}I error rate. The power depends on the reliability of a sample, which is represented by the standard error of the sample mean $ \sqrt{ \frac{\sigma^2}{M}} $, where $ \sigma^2$ is the unbiased estimate of the population variance and $M$ is the sample size \cite{helie2007understanding}. Therefore, if the threshold of the power is changed, the required sample size will also change. The power also depends on the threshold of the p-value, which is the significance level of the test and corresponds to the Type I error rate. This means that if the threshold of the power is fixed, the significance level affects the required sample size, that is, the number of measurements.

%the power depends not only on the sample size but also on the significance level, which corresponds to the Type I error rate, the significance level also affects the required sample size. This means that testing accuracy depends on the sample size, that is, the number of measurements. 
%Although these four characteristics affect the efficiency of locating bugs, they have not been mentioned in previous studies.

\section{Proposed Method}\label{proposed}
The proposed method consists of four approaches {\it cost-based binary search}, {\it early determination}, {\it finalization}, and {\it looking back}.

\subsection{Cost-based Binary Search}\label{sec-tree}
Binary search is an efficient array-search algorithm \cite{williams1976modification}, in which the location of the target value is recursively narrowed down by comparing the middle element of the array and target value. By selecting the center element of the array as the middle element, the expected search costs of the left and right subarrays are equal. The entire expected search cost is then minimized. 
Binary search is also effective in locating buggy segments in quantum programs. However, as [C1] states, the cost of the test depends on the position of the segment. Therefore, we use cost-based binary search in which the middle element is selected on the basis of the testing cost.
In our prior study \cite{sato2024locating}, the middle element $s_x$ was selected so that the highest total testing costs for searching the left and right sequences are as similar as possible. However, the expected search cost was not minimized in that way. Therefore, we update how to select the middle element.

Let $S_{l}$ be a sequence of segments of length $l$. If a segment $s_x$ ($1 \leq x \leq l-1$) is the middle element, the segment sequences from $s_1$ to $s_x$ and from $s_{x+1}$ to $s_l$ are called the left sequence and right sequence in terms of $s_x$, respectively. Since computational resources are expended for each gate execution, we define the testing cost $c_x$ as the number of quantum gates to be executed in the test of segment $s_x$. That is, $c_x = \sum_{i=1}^{x} g_i$, where $g_i$ denotes the number of quantum gates in $s_i$.

Let $s_{y}$ be the next segment to be tested in the left sequence $(1 \leq y \leq x-1)$. Note that $s_x$ is included in the left sequence, but it will not be tested. The expected testing cost of $s_{y}$ is the average of the testing costs from $s_1$ to $s_{x-1}$. This can be expressed by $ \frac{1}{x-1} \sum_{i=1}^{x-1} c_i$. From the test results of $s_{y}$, the segment sequence to be searched next is narrowed down to either $s_1$ to $s_y$ or $s_{y+1}$ to $s_x$. Thus, the length of the segment sequence to be searched after testing $s_{y}$ is expected to be $(y + (x - y))/2 = x/2$. The search continues until the length of the segment sequence reaches $1$, that is, the buggy segment is located. Therefore, the expected number of testing to locate the buggy segment is $\log_2 x$. This corresponds to the expected depth of the search path in the left sequence. The expected search cost of the left sequence $ec(s_x)_{left}$ until the buggy segment is located is calculated by multiplying the expected testing cost of each segment by the expected depth of the search path, that is, $ec(s_x)_{left} = (\frac{1}{x-1} \sum_{i=1}^{x-1} c_i)(\log_2 x)$. 

If there is a buggy segment in the right sequence of length $l-x$, the expected search cost $ec(s_x)_{right}$ can be similarly calculated as $ec(s_x)_{right} = (\frac{1}{l-x-1} \sum_{i=x+1}^{l-1} c_i)(\log_2 (l-x))$. Since we assume that a bug was detected in the entire sequence $S_{l}$ and we want to locate a segment that has the bug, if the bug is not detected in the test of segment $s_{l-1}$, we can conclude that the bug is contained in $s_{l}$. Therefore, the testing cost $c_l$ is not included in the calculation of $ec(s_x)_{right}$. In other words, since it was confirmed that there is a bug in $S_l$ by the test of $s_l$ before the search, we do not need to test $s_{l}$ again in the search.

Assumed that there is no bias in the probability that each segment contains a bug. The probability that the left sequence has a bug and is to be searched is $x/l$. Similarly, the probability that the right sequence is to be searched is $(l-x)/l$. Therefore, the expected search cost after the testing of segment $s_x$ is represented by $ec(s_x)_{left}*(x/l) + ec(s_x)_{right}*((l-x)/l)$. Adding the testing cost $c_x$ of $s_x$ to this formula results in the expected search cost $ec(s_x)$ when $s_x$ is selected as the middle element.

%The numbers of search paths in the left and right sequences are equal to the numbers of the segments, that is, $x$ and $l - x$, respectively. 
%Finally, the expected search cost $ec(s_x)$ when $s_x$ is selected as the middle element is expressed with the following formula.
\begin{align}
\label{fec}
ec(s_x) = & ec(s_x)_{left}*(x/l) + ec(s_x)_{right}*((l-x)/l) + c_x \nonumber \\
 = & \left( \frac{1}{x-1} \sum_{i=1}^{x-1} c_i \right)(\log_2 x)(x/l) \nonumber \\
    & + \left( \frac{1}{(l-x)-1} \sum_{i=x+1}^{l-1} c_i \right)(\log_2 (l-x))((l-x)/l) \nonumber \\
   & + c_x
\end{align}
\begin{comment}
ec(s_x) = & \left( ec_{left}(x) + ec_{right}(l-x) \right) / l + c_x \nonumber \\
 = & \left(
     \left( \frac{1}{x-1} \sum_{i=1}^{x-1} c_i \right)(\log_2 x)(x) \right. \nonumber \\
    & \left. + \left( \frac{1}{(l-x)-1} \sum_{i=x+1}^{l-1} c_i \right)(\log_2 (l-x))(l-x)
    \right)
    / l \nonumber \\
   & + c_x
\end{comment}

The middle element $s_x$ is selected so that the expected search cost is minimized. Therefore, the index $x$ of middle element $s_x$ is given as
\begin{eqnarray*}
\label{fmid}
\mathop{\arg \min}_{1 \leq x \leq l-1} ec(s_x).
\end{eqnarray*}

\begin{figure*}[hbt]
\centering
\scalebox{0.4}{\includegraphics[bb=0 0 1246 356]{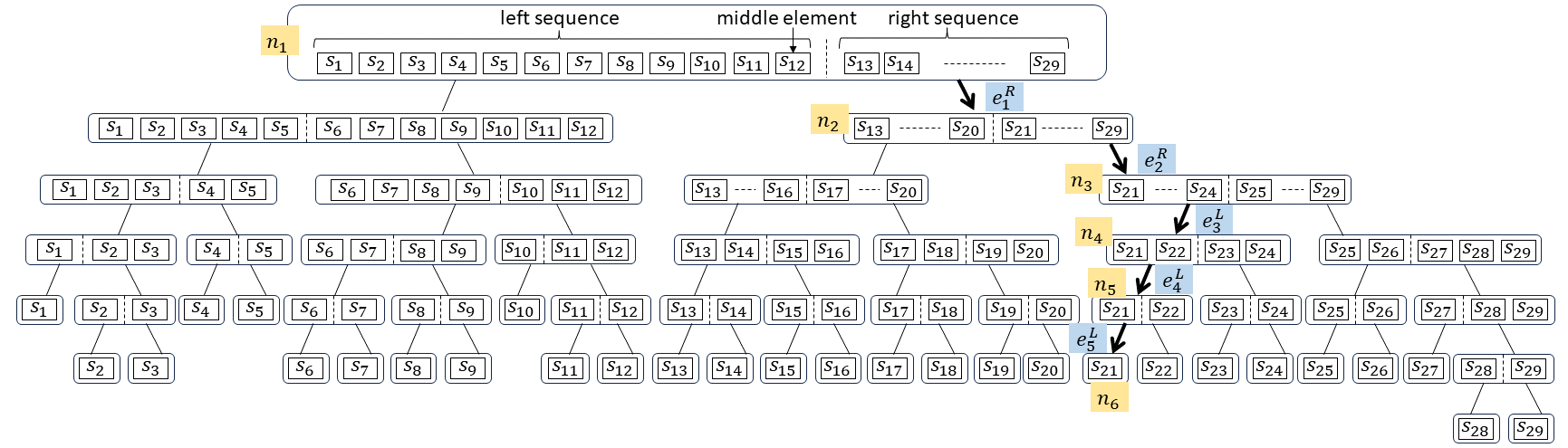}}
\caption{Example of cost-based binary search tree}
\label{tree_r1}
\end{figure*}
Figure \ref{tree_r1} shows the cost-based binary search tree of the quantum program shown in Figure \ref{segmentation}. A node in the tree is associated with a segment sequence to be searched, which we call a target sequence. The dashed line in a target sequence indicates the output state of the middle element to be tested. When testing a target sequence, all segments ahead of the middle element will be executed, including those not included in the target sequence. The sequence of segments ahead of the middle element is called an executed sequence. If a bug is detected from the test at node $n_i$, move to the left child node with the edge $e_i^L$. Otherwise, transit to the right child node with the edge $e_i^R$.
At node $n_4$ in Figure \ref{tree_r1}, the target sequence consists of $s_{21}$, $s_{22}$, $s_{23}$, and $s_{24}$. The executed sequence is from $s_1$ to $s_{22}$. If a bug is detected in the test at $n_4$, the search moves to a node consisting of $s_{21}$ and $s_{22}$, which is labeled as $n_5$. If not, it moves to a node consisting of $s_{23}$ and $s_{24}$.

\IncMargin{1em}
\begin{algorithm}
\setstretch{0.9} %ここの数値で行間を調整
\DontPrintSemicolon

\SetKwInOut{Input}{input}
\SetKwInOut{Output}{output}
\SetKwFunction{FcomposeNode}{composeNode}
\SetKwProg{Fn}{def}{\string:}{}

\Input{$searched\_quantum\_program$}
\Output{$root\_node$ : indicating the search tree}
\BlankLine

\Fn(){\FcomposeNode{$cur\_node$}} {
    $target\_seq$ $ \leftarrow $ $cur\_node.target\_seq$ \; \label{ln10}
    \If{the length of $target\_seq > 1$} { \label{ln09}
        $ec\_list$ $ \leftarrow $ [] \;
        \For{$s_x$ $ \in $ $target\_seq$} {
            $ec\_list$.append(ec($s_x$)) \; \label{ln03}
        }
        $cur\_node.middle\_el$ $ \leftarrow $ argmin($ex\_list$) \; \label{ln04}
        $cur\_node.left\_node$ $ \leftarrow $ Node($target\_seq$[:$middle\_el$+1]) \; \label{ln05}
        $cur\_node.right\_node$ $ \leftarrow $ Node($target\_seq$[$middle\_el$+1:]) \; \label{ln06}
        composeNode($cur\_node.left\_node$) \; \label{ln07}
        composeNode($cur\_node.right\_node$) \; \label{ln08}
    }
}
\BlankLine
$root\_node$ $ \leftarrow $ Node($searched\_quantum\_program$) \; \label{ln01}
composeNode($root\_node$) \; \label{ln02}
\Return $root\_node$ \;
\caption{Algorithm to compose cost-based binary search tree}
\label{algorithm01}
\end{algorithm}\DecMargin{1em}
%ルートノードから探索木を構成していく。$searched\_quantum\_program$は、バグを特定する対象の、セグメント分けされたプログラムである。Line \ref{}では、$searched\_quantum\_program$をtarget sequenceとして持つノードを作成している。これをルートノードとして、サブルーチン$composeNode$を呼び出す（line \ref{ln02}）。$composeNode$は、引数として受け取った$current\_node$に対して、left child nodeとright child nodeを作成する。ただし、$current\_node$に対応付けられたtarget sequenceの長さが1の場合は、そのノードはリーフノードとなるため、child nodesは作成しない（line \ref{ln09}）。line \ref{ln03}では、Formula \ref{fec}に従って、expected search cost ec(s_x)をそれぞれのsegmentに対して計算する。そして、Formula \ref{fmd}に示したとおり、expected search costが最小となるセグメントを、middle segmentとする（line \ref{ln04}）。次にline \ref{ln05}と\ref{ln06}ではそれぞれ、left sequenceとrightsequenceを対応付けて、left child nodeとright child nodeを作成している。line \ref{ln07}と\ref{ln08}では、作成したleft child nodeとright child nodeをに対して、サブルーチン$composeNode$を再帰的に呼び出している。
The detailed algorithm to compose a search tree is shown in Algorithm \ref{algorithm01}.
The search tree is constructed from the root node. This algorithm receives the $searched\_quantum\_program$ as input, which is the segmented program to be searched. In line \ref{ln01}, a node that has $searched\_quantum\_program$ as the target sequence is created. With this as the root node, a subroutine $composeNode$ is called in line \ref{ln02}, which receives an argument $cur\_node$. This subroutine creates left and right child nodes of $cur\_node$ by selecting the middle element in accordance with Formula \ref{fec}. 
In line \ref{ln10}, the target sequence in $cur\_node$ is assigned to $target\_seq$. In line \ref{ln03}, the expected search cost ec($s_x$) is calculated for each segment consisting of $target\_seq$. In line \ref{ln04}, the segment with the lowest expected search cost is selected as the middle segment, and its index is assigned to $cur\_node.middle\_el$. In lines \ref{ln05} and \ref{ln06}, the left and right child nodes that have the left sequence and right sequence, respectively, are created. Subroutine $composeNode$ is recursively called for these left and right child nodes in lines \ref{ln07} and \ref{ln08}, respectively. If the length of a child node is $1$, it is a leaf node. In that case, $composeNode$ is no longer called recursively (line \ref{ln09}).

\subsection{{\bf Early Determination}}\label{early}
In accordance with the cost-based binary search tree, we search for a buggy segment. On the basis of [C4], we introduce an approach to reduce the search cost, which is called early determination. This is based on the assumption that statistically sufficient accuracy is over-performance for narrowing down (not determining) a buggy segment, and it may be more efficient to reduce the number of measurements by taking the risk of return in the search. The probability of return is evaluated in Section \ref{risk} on the basis of Bayes' theorem. 
%It also shows that the risk of return becomes smaller in the search if a certain condition is satisfied. Ohterwise, the risk is reduced by looking back described in Section \ref{suspicious}.

As another motivation of early determination, we focus on the ``reinforcement" relation between determinations, which is based on [C3].
%A search path from node $n_1$ to $n_k$ can be denoted as a sequence of edges $[e_1^{d_1}, ... , e_i^{d_i}, e_{i+1}^{d_{i+1}}, ... , e_{k-1}^{d_{k-1}}]$. 
Let $S_i$ be the executed sequence of $n_i$, which is a node included in a search path from $n_1$ to $n_k$ ($1 \leq i \leq k-1$). The sequence of edges from node $n_1$ to $n_k$ is denoted as $[e_1^{d_1}, ... , e_i^{d_i}, e_{i+1}^{d_{i+1}}, ... , e_{k-1}^{d_{k-1}}]$, where $L$ or $R$ is assigned to $d_i$.
Assume that there is no bug in the executed sequence $S_i$, which corresponds to the null hypothesis of the test at $n_i$. The Type I error rate of the test at the node is denoted as $ \alpha $. If we determine at $n_i$ that there is a bug in $S_i$, that is $d_i=L$, the probability of making this determination under the null hypothesis is $ \alpha $. In accordance with the structure of the search tree, the executed sequence $S_{i+1}$ of $n_{i+1}$ is included in the executed sequence $S_i$ of $n_i$. This means that there is also no bug in $S_{i+1}$ under the null hypothesis. Therefore, the probability of determining $d_{i+1}=L$ at $n_{i+1}$ is also $ \alpha $ under the null hypothesis of $n_i$. Finally, the probability of determining $d_i=d_{i+1}=L$ is $ \alpha ^2 $. This means that by proceeding from $n_i$ to $n_{i+1}$ with $d_i=d_{i+1}=L$, the null hypothesis of $n_i$ can be rejected with more certainty at $n_{i+1}$ than at $n_i$. More certain rejection of the null hypothesis of $n_i$ corresponds to upholding the past determination at $n_i$. We call this relation $e_{i+1}^L$ reinforces $e_i^L$. This reinforcement relation generally holds not only for $e_i^L$ and $e_{i+1}^L$ but also for any $e_{j}^L$ and $e_i^L$ ($i < j$).
The same is applied for $d_i = d_j = R (i < j)$. In that case, $e_{j}^R$ reinforces $e_i^R$. 
In the example in Figure \ref{tree_r1}, $e_2^R$ reinforces $e_1^R$, $e_4^L$ and $e_5^L$ reinforce $e_3^L$, and $e_5^L$ also reinforces $e_4^L$.

The fact that past determinations may be confirmed by later determinations motivates early determination. However, an incorrect determination is not reinforced later; thus, we need other approaches to modify the incorrect determination, which are described in Sections \ref{finalization} and \ref{suspicious}.

\subsection{{\bf Finalization}}\label{finalization}
Early determination is based on the assumption that sufficient accuracy is not necessary when advancing the search. When finally locating a buggy segment, however, the test should be conducted with sufficient accuracy.
Therefore, on the basis of [C2], the proposed method executes finalization when the binary search reaches a leaf node and locates the buggy segment $s_x$. Finalization consists of the tests of $s_{x-1}$ and $s_x$ with sufficient accuracy. It should be confirmed that the segments from $s_1$ to $s_{x-1}$ do not contain a bug by the test of $s_{x-1}$, and that the segments from $s_1$ to $s_x$ contain a bug by the test of $s_x$, with sufficient accuracy. If finalization reveals an incorrect determination at a node, the search will return to that node. 
In the example in Figure \ref{tree_r1}, when the search reaches leaf node $n_6$, the tests at $n_2$ and $n_5$ are conducted with sufficient accuracy as finalization. That is, measurements of segments $s_{20}$ and $s_{21}$ are added to sufficiently confirm that there are no bugs from $s_1$ to $s_{20}$ and there is a bug from $s_1$ to $s_{21}$, respectively. For example, if finalization reveals that there is a bug from $s_1$ to $s_{20}$, that is, the determination at $n_2$ is incorrect, the search returns to $n_2$.

\subsection{{\bf Looking Back}}\label{suspicious}
In addition to finalization, we introduce looking back to modify incorrect determinations. First, we show that we only need to focus on the last $L$ edge in a search path if we have incorrectly determined that there is a bug. Assume that the binary search is executed from $n_1$ to $n_k$ with the path $[e_1^{d_1}, ..., e_h^L, ..., e_i^L, e_{i+1}^R, ..., e_{k-1}^R]$ in which $e_h^L$ is an arbitrary $L$ edge from $n_1$ to $n_i$ ($1 \leq h \leq i-1$) and $e_i^L$ is the last edge of $L$. 
Since the executed sequence $S_i$ of $n_i$ is included in the executed sequence $S_h$ of $n_h$, if there is no bug in $S_h$, there is also no bug in $S_i$. That is, if $e_h^L$ is incorrect, $e_i^L$ is also incorrect. Therefore, when we want to confirm that there is an incorrect $L$ edge in the path from $n_1$ to $n_i$, we only need to check whether the last $L$ edge $e_i^L$ is incorrect.

Next, we focus on the successor $e_{i+1}^R, ..., e_{k-1}^R$. If this is long, it suggests that $e_i^L$ is likely to be incorrect. If $e_i^L$ is incorrect, that is, $S_i$ does not include a bug, the executed sequences $S_{i+1}, ..., S_{k-1}$ also do not include a bug. In that case, $R$ edges appear in succession if the determinations are correctly executed.
Therefore, if $R$ edges appear more than a certain number of times in succession, we should suspect that $e_i^L$ is incorrect. Looking back confirms the determination of the last $L$ edge by the test with sufficient accuracy. In this case, $n_i$ and $e_i^L$ are called the suspicious node and edge, respectively. If the determination turns out to be incorrect, the search returns to the suspicious node. 

The same is applied for the path $[e_1^{d_1}, ..., e_h^R, ..., e_i^R, e_{i+1}^L, ..., e_{k-1}^L]$ in which $e_h^R$ is an arbitrary $R$ edge from $n_1$ to $n_i$ ($1 \leq h \leq i-1$) and $e_i^R$ is the last $R$ edge. Since the executed sequence $S_h$ of $n_h$ is included in the executed sequence $S_i$ of $n_i$, if there is a bug in $S_h$, there is also a bug in $S_i$. Therefore, we can focus on the correctness of $e_i^R$. If the successor $e_{i+1}^L, ..., e_{k-1}^L$ is long, it should be suspected that there is a bug in $S_i$, and the last $R$ edge, $e_i^R$, should be confirmed by the test with sufficient accuracy. In this case, $n_i$ and $e_i^R$ are called the suspicious node and edge, respectively.
%Similarly, if $L$ edges appear in succession, the last $R$ edge is looked back to.

In the example of Figure \ref{tree_r1}, the executed sequence of $n_1$ is segment $s_1$ to $s_{12}$. They are included in the executed sequence of $n_2$, that is, $s_1$ to $s_{20}$. If there is a buggy segment from $s_1$ to $s_{12}$, it is also included from $s_1$ to $s_{20}$. Therefore, we should check $e_2^R$ to confirm whether there is an incorrect $R$ edge from $n_1$ to $n_3$. Let $D$ be the threshold of successive occurrence of $L$ edges for looking back. When $D$ is defined as 3, since $L$ edges of $e_3^L$, $e_4^L$, and $e_5^L$ appear in succession, the last $R$ edge, $e_2^R$, is suspicious and should be looked back to. %The theoretical meaning of looking back is referred to in Section \ref{risk}.
%In the example of Figure \ref{tree}, since $e_3^R$, $e_4^R$, and $e_5^R$ appear $D=3$ times in succession, we look back to the last $L$ edge, $e_2^L$, and test it with sufficient accuracy.
%Similarly, if $L$ edges appear in succession, the last $R$ edge is looked back to.

\subsection{Search Algorithm}\label{sec_algo}
The detailed search algorithm based on the approaches described in Sections \ref{early}, \ref{finalization}, and \ref{suspicious} is shown in Algorithm \ref{algorithm02}.

\IncMargin{1em}
\begin{algorithm}
\setstretch{0.9} %ここの数値で行間を調整
\DontPrintSemicolon

 \SetKwInOut{Input}{input}
 \SetKwInOut{Output}{output}

\Input{$tree$%, $m_{unit}$, $M_{max}$, $D$, $Sig$, $T\_power$, $T\_final\_right$, $T\_early\_left$, $T\_early\_right$
}
\Output{$cur\_node$ : indicating the buggy segment}
\BlankLine

initialize($tree$, 'Undetermined')\; \label{al00}
\While{True} {
  $tested\_node$ $ \leftarrow $ get\_suspicious\_node($tree$, D)\; \label{al01}
  \If{$tested\_node$ is not None} { \label{al_add01}
    \If{$tested\_node.dtmn$ is LeftFinalized or RightFinalized} { \label{al_add02}
      $tested\_node$ $ \leftarrow $ None \; \label{al_add03}
    }
  }
  \If{tested\_node is None}{ \label{al02}
    $cur\_node$ $ \leftarrow $ get\_deepest\_reachable\_node($tree$)\; \label{al03}
    \uIf{$cur\_node$ is a leaf node}{ \label{al04}
      $node\_for\_input$ $ \leftarrow $ get\_input($cur\_node$)\; \label{al05}
      $node\_for\_output$ $ \leftarrow $ get\_output($cur\_node$)\; \label{al06}
      \uIf{$node\_for\_input$ is finalized $ \land $ $node\_for\_output$ is finalized}{ \label{al07}
%       \State buggy\_segment \leftarrow cur\_node.segment \label{al08}
        break\; \label{al09}
      }
      \uElseIf{$node\_for\_input$ is not finalized}{ \label{al10}
        $tested\_node$ $ \leftarrow $ $node\_for\_input$ \; \label{al11}
      }
      \Else{ \label{al12}
        $tested\_node$ $ \leftarrow $ $node\_for\_output$ \; \label{al13}
      } \label{al14}
    }
    \Else{ \label{al15}
      $tested\_node$ $ \leftarrow $ $cur\_node$ \; \label{al16}
    }
  } \label{al17}
  \uIf{$tested\_node.num\_m$ $<$ $M_{max}$} { \label{ad_add04}
    \If{($M_{max} - tested\_node.num\_m$) $ \geq $ $m_{unit}$} { \label{al_add05}
      measure($tested\_node$, $m_{unit}$)\; \label{al18}
    }
    \Else {
      measure($tested\_node$, $M_{max} - tested\_node.num\_m$)\; 
    }
  } \label{al_add06}
  \Else{ \label{al_add07}
    \Return None \; \label{al_add08}
  }
  $p \text{-} value$, $power$ $ \leftarrow $ calc\_indicator($tested\_node$)\; \label{al19}
  \uIf{$p \text{-} value$ $ \leq $ Sig $ \land $ power $ \geq $ T\_power}{ \label{al20}
    $tested\_node.dtmn$ $ \leftarrow $ 'LeftFinalized'\; \label{al21}
  }
  \uElseIf{$p \text{-} value$ $ \geq $ T\_upper\_p}{ \label{al22}
    $tested\_node.dtmn$ $ \leftarrow $ 'RightFinalized'\; \label{al23}
  }
  \uElseIf{$p \text{-} value$ $ \leq $ Sig\_relaxed $ \land $ $power$ $ \geq $ T\_power\_relaxed}{ \label{al24}
    $tested\_node.dtmn$ $ \leftarrow $ 'LeftEarly'\; \label{al25}
  }
  \uElseIf{$p \text{-} value$ $ \geq $ T\_upper\_p\_relaxed}{ \label{al26}
    $tested\_node.dtmn$ $ \leftarrow $ 'RightEarly'\; \label{al27}
  }
  \Else{ \label{al28}
    $tested\_node.dtmn$ $ \leftarrow $ 'Undetermined'\; \label{al29}
  } \label{al30}
} \label{al31}
\Return $cur\_node$ \; \label{al32}

\caption{Search Algorithm}\label{algorithm02} 
\end{algorithm}\DecMargin{1em}

The input parameter $tree$ is an object corresponding to the search tree created with Algorithm \ref{algorithm01}. As shown in Figure \ref{tree_r1}, the search tree consists of nodes and edges. Each node has an attribute $dtmn$ that represents the determination of the test at the node. The value of $dtmn$ is `LeftFinalized,' `RightFinalized,' `LeftEarly,' `RightEarly,' or `Undetermined,' respectively meaning `determined as buggy with sufficient accuracy,' `determined as bug-free with sufficient accuracy,' `determined as buggy by early determination,' `determined as bug-free by early determination,' and `undetermined.' In line \ref{al00}, attribute $dtmn$ is initialized to `Undetermined' at all nodes in $tree$. The procedure get\_suspicious\_node in line \ref{al01} checks whether there is a suspicious node in looking back, that is, there are $D$ or more successive edges with the same label on the current search path. If the successive edges are found, the suspicious node is assigned to $tested\_node$ for looking back. Otherwise, None is substituted. If the determination at the suspicious node has already been made with sufficient accuracy, no further measurements are added (line \ref{al_add02}). In that case, $tested\_node$ is cleared to None (line \ref{al_add03}).

The search of a binary search tree proceeds from the upper node to the lower node. However, the determination at a higher node may be changed by looking back. This results in the search path from the root node is also changing. Therefore, in line \ref{al03}, the procedure get\_deepest\_reachable\_node searches the tree and updates the current node by the deepest node reachable from the root node. If the updated current node is a leaf, finalization is executed. To test the input and output quantum states of the located buggy segment, $node\_for\_input$ and $node\_for\_output$ are obtained, respectively  (lines \ref{al04} to \ref{al06}). In line \ref{al10}, it is checked if the determination of those nodes are made with sufficient accuracy. If so, since it means that the buggy segment is finally located, the procedure goes from lines \ref{al09} to \ref{al32} and terminates after returning $cur\_node$, which indicates the located buggy segment. Otherwise, a node the determination of which has not been finalized is assigned to $tested\_node$ (lines \ref{al11} and \ref{al13}). If the current node is not a leaf, the current node will be $tested\_node$ (lines \ref{al15} and \ref{al16}).

Next, the test corresponding to $tested\_node$ is executed. Node $tested\_node$ is implemented as an object and has an attribute $num\_m$ that represents the number of the past measurements. The measurement results of the past $num\_m$ times at the node is also saved in that object. The constants $m_{unit}$ and $M_{max}$ represent the number of measurements per unit and maximum number of measurements, respectively. Until $num\_m$ exceeds $M_{max}$, $m_{unit}$ times measurements are additionally executed (lines \ref{ad_add04} to \ref{al_add06}). Note that $num\_m$ is updated to $num\_m + m_{unit}$ in the function $measure$. If $num\_m$ reaches $M_{max}$ before the determination is finalized, the search fails (line \ref{al_add08}). 
As described in Section \ref{chara}, we used the chi-square test for testing a segment, which statistically compares the results of Z-basis measurements with its oracle. In line \ref{al19}, the p-value and power of the test are calculated from the measurement results. The oracle of the measurement results is derived from the expected amplitudes of the tested quantum state. The null hypothesis for the chi-square test is that there is no bug in the executed sequence. In lines \ref{al20} to \ref{al30}, the $dtmn$ at $tested\_node$ is updated on the basis of the calculated p-value and power. The $Sig$ is the significance level, which is the threshold of the p-value, and $T\_power$ is the threshold of the power. If $p \text{-} value \leq Sig \land power \geq T\_power$ holds, it is determined that there is a bug with sufficient accuracy. In that case, `LeftFinalized' is assigned to $dtmn$ in line \ref{al21}. In line \ref{al22}, $T\_upper\_p$ is used as the threshold of the p-value for determining the absence of a bug, and is set to a value close to 1 \cite{huang2019statistical}. Since the power represents the probability of correct determination when the alternative hypothesis holds, i.e., when there is a bug,  it is not used as an indicator when determining the absence of a bug. The $Sig\_relaxed$ and $T\_power\_relaxed$ are the thresholds of the p-value and power when determining the presence of a bug in early determination, respectively. Similarly, $T\_upper\_p\_relaxed$ is the threshold of the p-value referred to when determining the absence of a bug in early determination. By changing these thresholds, we can adjust how early the early decision is made.

\section{Experiments}\label{exp}
\subsection{Comparison with Naive Methods}
%コストでは勝っているが、Standard deviationがたまに負けることがあることの考察　→平均コストが最小になるようにしているので、最良ケースと最悪ケースの差が必ずしも小さいとは限らない。そのため、ばらつきが大きくなることもある。よって、pos13への改良後は、Standard deviationは目安に過ぎない。pos5の時は、どこにバグある場合でもコストが平準化されるようにしていたため、Standard deviationは小さくなる傾向があった。
We implemented our proposed method and applied it to arbitrarily generated quantum programs of several different sizes. We specified the numbers of qubits, segments, and quantum gates. Generated quantum programs consist of arbitrary single- and two-qubit gates and satisfy those conditions. The number of quantum gates in a segment differs for each segment. For each combination of the conditions, 1000 quantum programs were generated, and the proposed method was applied to each of them.

In each program, a buggy segment was made by arbitrarily replacing one quantum gate in a segment with another.% so that the output state of the buggy segment changes enough to be detected by statistical test. 
%今回はバグが各セグメントに存在する確率は均一と仮定して、ランダムに選択したセグメントにbugをinjectした。
Assuming that there is no bias in the probability that each segment contains a bug, the buggy segment was randomly selected. 
For each of the 1000 quantum programs, we evaluated the probability of successfully locating a buggy segment and average search cost. Two types of search costs were calculated. One is the average cost for the cases of correctly locating the buggy segment, and the other is the average cost for both cases of success and failure to locate the buggy segment. The search cost is calculated from the number of quantum gates executed in the search. When calculating the cost, a single- and two-qubit gate are not distinguished. In an actual quantum computer, only native gates can be executed, which are converted from other (non-native) quantum gates. The type of native gates depends on the device of the quantum computer. The testing cost of a segment depends on the number of native gates consisting of that segment. For sake of simplicity, we assume that all single- and two-qubit gates can be executed and calculate the testing cost as the number of those gates.

In cost-based binary search in our prior study \cite{sato2024locating}, the middle element was selected so that the highest search cost of the left and right sequences are equalized. Therefore, the standard deviation of the cost was also used as an evaluation criterion. In this paper, however, since the middle element is selected so that the expected search cost is minimized, the standard deviation of the cost is not expected to decrease. Therefore, in this experiment, it was not used as an evaluation criterion.

%As mentioned in Section \ref{sec_algo}, we used the chi-square test to test the output state of the middle segment. In the chi-square test, the estimated absolute squares of the amplitudes are represented by the actual results of measurements. They are compared to the oracles calculated from the expected amplitudes.
%expected measurement resultが5よりも少ない基底がある場合は、Yates's correction for continuityを適用する。
In the chi-squared test, if there is a basis with less than five measurements, Yates's correction for continuity is applied.
%the absolute square of the amplitude is compared with its oracle from the results of Z-basis measurements when testing the output state of the middle segment. 
%測定は100回ずつ行う。まずテスト対象の出力状態を100回測定した後、その100回の測定結果に対してカイ二乗テストを実施する。その結果、p-valueなどのaccuracy indicatorsが閾値を超えず、バグの有無を判定できなかった場合は、さらに100回の測定を実行し、合計200回の測定結果に基づいてカイ二乗テストを実施する。これをaccuracy indicatorsが閾値を超えるか、または測定回数が上限に達するまで繰り返す。The upper limit of number of measurements is $2^{n*10000}$, where $n$ is the number of qubits. 測定回数が上限に達した場合は、そこでバグ位置特定失敗として、探索を終了する。
%In the chi-square test, the p-value and power of the test are referred to as accuracy indicators. 
For determining the presence of a bug with sufficient accuracy, the threshold of the p-value $Sig$ and the threshold of the power $T\_power$ are set to 0.05 and 0.8, respectively. For early determination, the relaxed threshold $Sig\_relaxed$ and $T\_power\_relaxed$ are defined as 0.1 and 0, respectively, That is, the power is not referred to in early determination. When determining the absence of a bug with sufficient accuracy, the threshold of the p-value $T\_upper\_p$ is set to 0.8, and it is relaxed to 0.6 in early determination. %The power is not referred to when determining the absence of a bug because it is similar to the significance level when there is no bug. 
Looking back is executed when the same kind of edge ($R$ or $L$ edge) appears three times in succession, that is, $D=3$. %1ユニットの測定回数$m_{unit}$と測定の上限回数$M_{max}$は、それぞれ$100$、$2^{n}*10000$とする, where $n$ is the number of qubits。カイ二乗テストで比較するthe amplitudeの数は$2^{n}$であるため、それに比例して測定回数の上限も増加させる。
The number of measurements per unit $m_{unit}$ and the maximum number of measurements $M_{max}$ are $100$ and $100,000$, respectively.
% where $n$ is the number of qubits. Since the number of bases which corresponds to the categories compared in the chi-square test is $2^{n}$, we increase the maximum number of measurements which corresponds to the maximum number of samples proportionally.

We also implemented naive linear search and naive (non-cost-based) binary search, in which the central segment is selected as the middle element, as comparison methods. Our approaches described in Section \ref{proposed} are not applied to these comparison methods. 
With these comparison methods, the chi-square test is also used for testing segments with the same thresholds as with the proposed method with sufficient accuracy. That is, the thresholds of the p-value and power are 0.05 and 0.8, respectively, for determining the presence of a bug. When determining the absence of a bug, the threshold of the p-value is 0.8. 
As with the proposed method, %測定は$m_{unit}=100$回ごとにカイ二乗テストを行い、p-value and powerが閾値を超えるかを確認する。これらが閾値を超えずに、測定回数が上限$M_{max}=2^{n}*10000$に達した場合、そこで探索失敗とする。
the chi-square test is executed every $m_{unit}=100$ measurements to check whether the p-value and power exceed those thresholds. If the number of measurements reaches $M_{max}$ without exceeding the thresholds, the search fails. 

%In the comparison methods, we use the chi-square test to test each segment as in the proposed method, 
If the output state of the middle segment does not change significantly before and after making a buggy segment, it is difficult to detect the bug from the chi-square test. In such a case, neither the proposed method nor the comparison methods can locate the buggy segment; thus, a comparative evaluation is not possible. 
%Our method is efficient for locating bugs based on statistical testing. If the statistical testing cannot correctly determine the presence or absence of bugs with a certain degree of accuracy, it is difficult to comaratively evaluate the proposed method to naive methods.
Therefore, if it is difficult to detect a bug in a generated quantum program with the chi-square test, the program is excluded from the experiment. More specifically, we calculated the difference in the absolute square of the amplitude for each basis before and after making the buggy segment. If the sum of these differences was 0.05 or less, the quantum program was excluded.

The experiment was conducted through simulation on a classical computer using Qiskit\textregistered \cite{cross2018ibm}. The classical computer was Windows 10\textregistered Professional machine equipped with two Intel\textregistered \ Core\texttrademark \ i9-10900 2.80-GHz processors with 10 cores and 64-GB memory. %It also has an Intel\textregistered UHD Graphics 630 GPU.
The experimental results are listed in Table \ref{table01}. 
\begin{table*}
\centering
\scalebox{0.6}{\includegraphics[bb=-30 0 1591 300]{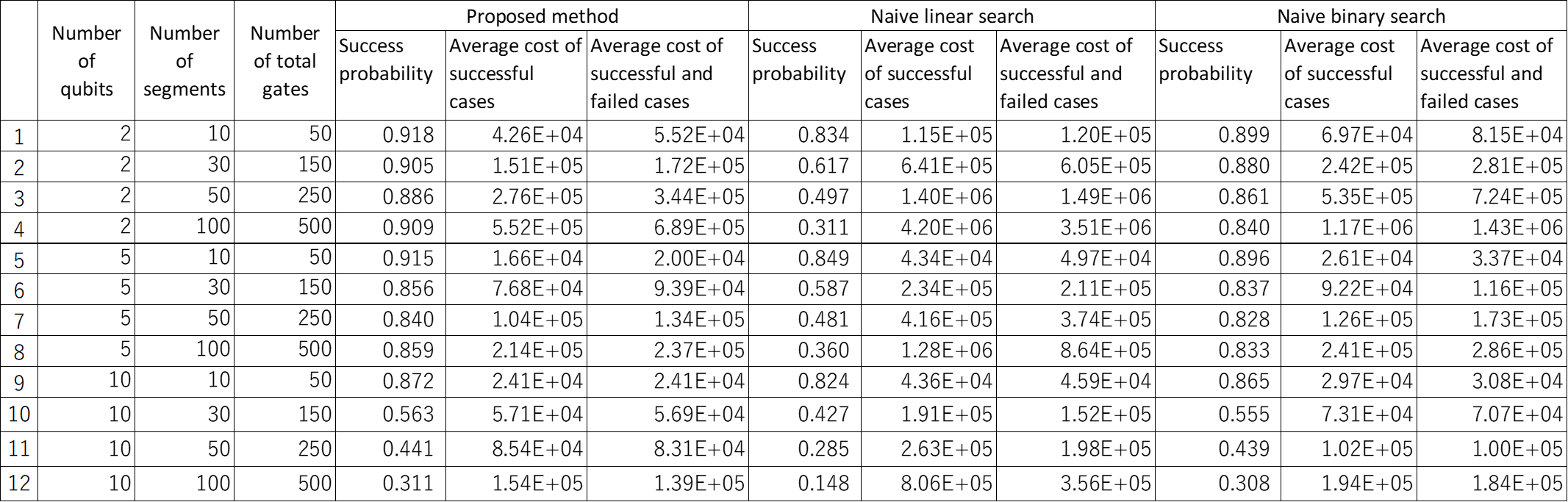}}
\caption{Experimental results}
\label{table01}
\end{table*}
These results indicate that the average costs of the proposed method are lower than those of the comparison methods. The results also indicate that the success probabilities of the proposed method are higher than those of the comparison methods. If the difference in the output state caused by a bug is small, it is difficult to determine the presence of the bug by the test with sufficient accuracy. Therefore, it is likely that the number of measurements reaches the upper limit and the search fails. Since the proposed method uses early determination, the search is more likely to proceed before the number of measurements reaches the upper limit than with the comparison method. That is, early determination also contributes to the improvement of the success probabilities. 
In cases where buggy segments could not be located, the search cost tends to be large because the number of measurements for the test of a segment reaches the maximum number when the search fails. Since the success probabilities of the comparison methods are lower than that of the proposed method, the average search costs of the comparison methods for both cases of success and failure tend to be larger. 
%This means that the fact that the average search cost of successful and failed cases in the proposed method is lower in Table \ref{table01} is not sufficient to show the efficiency of the proposed method
We also confirmed that the average search cost of only successful cases is also lower than those of the comparison methods. This means that the proposed method not only has higher success probabilities but is also more efficient than the comparison methods in successful cases.

%########################################################################
%初めからm_unitも量子ビット数に比例して増加させるか？→1セットの測定回数を(2^n)*100にしたところ、量子ビット数が２の場合に壊滅的な成功率になったためNG
Table \ref{table01} also shows that the success probability decreases as the number of qubits $n$ increases. The reason for the decrease is the insufficient number of measurements. The number of bases that corresponds to the categories of the compared distributions is $2^{n}$. 
%カイ二乗検定において、カテゴリ数は自由度としてp値やpower of the testの計算に使用される。サンプル数が一定の場合、カテゴリ数が大きくなるほど、p値は大きく、powerは小さくなりやすい。即ちテストの精度が低くなりやすい。そのため、量子ビット数$n$の増加に伴い、測定回数も増加させる必要がある。
In the chi-square test, the number of categories is used to calculate the p-value and power of the test. If the sample size is constant, the larger the number of categories, the larger the p-value and smaller the power \cite{helie2007understanding}. When there is a bug in a segment, the presence of the bug is determined by checking $p \text{-} value \leq Sig \land power \geq T\_power$, as shown in Algorithm \ref{algorithm02}. Therefore, if the number of categories is large, it is difficult to correctly determine the presence of the bug. %This means that testing accuracy is likely to be lower for the same sample size. 
If the number of the samples is increased, the p-value and power tend to decrease and increase, respectively.
This corresponds to the intuition that if the number of categories is large and there are few samples, sampling bias is likely to occur and we cannot be confident in the test result. 
Therefore, the larger the number of samples, the more correctly the presence of the bug is determined. The required number of measurements increases as the number of qubits $n$ increases. %On the other hand, since the absence of a bug is determined by $p \text{-} value \geq T\_upper\_p$, increase of the categories does not affect the determination of the absence of the bug
We conducted an additional experiment in which $m_{unit}$ and $M_{max}$ were increased from $100$ and $100,000$ to $10,000$ and $10,000,000$, respectively, for the number of qubits $n = 10$. Though the increased number of measurements increased the search costs, the success probabilities improved, as shown in Table \ref{table01_m_increased}. %This experiment was conducted for 100 quantum programs in each condition.
\begin{table}
\centering
\scalebox{0.55}{\includegraphics[bb=0 0 873 210]{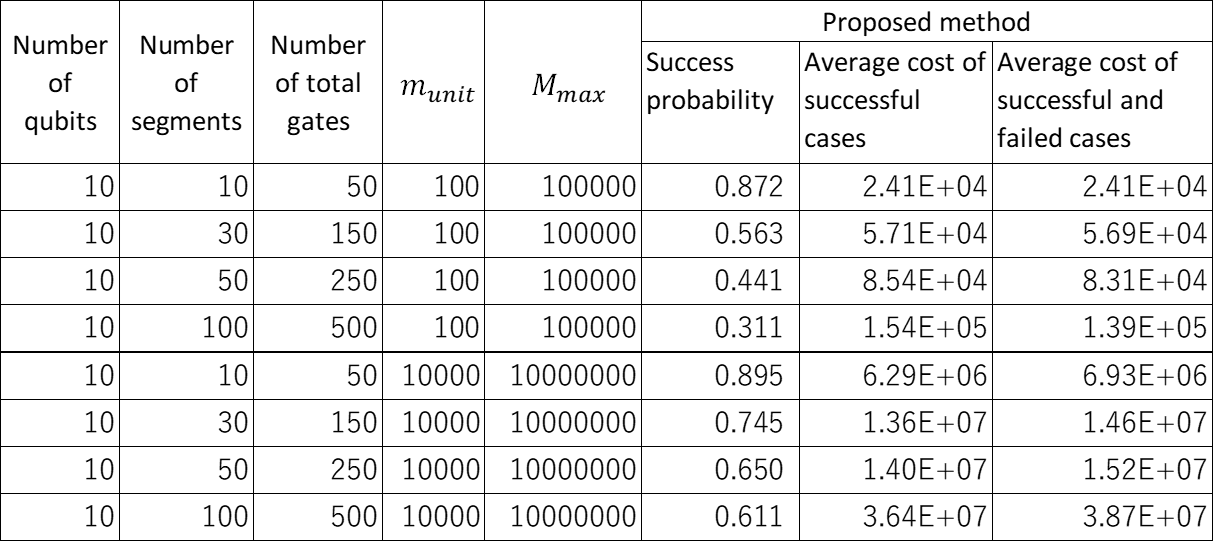}} %0 0 1786 205
\caption{Results before and after increasing measurements}
\label{table01_m_increased}
\end{table}

\subsection{Effectiveness of Each Approach}
To evaluate the effectiveness of each approach, we conducted the same experiments without applying each one. %The following experiments were also conducted for 100 quantum programs in each condition.

\subsubsection{Effectiveness of cost-based binary search}
The results of the experiment without applying cost-based binary search is shown in the left column of Table \ref{table02}. The search tree was changed from a cost-based binary search tree to a naive binary search tree. The other approaches, early determination, finalization, and looking back, were applied.
\begin{table*}
\centering
\scalebox{0.6}{\includegraphics[bb=-100 0 889 250]{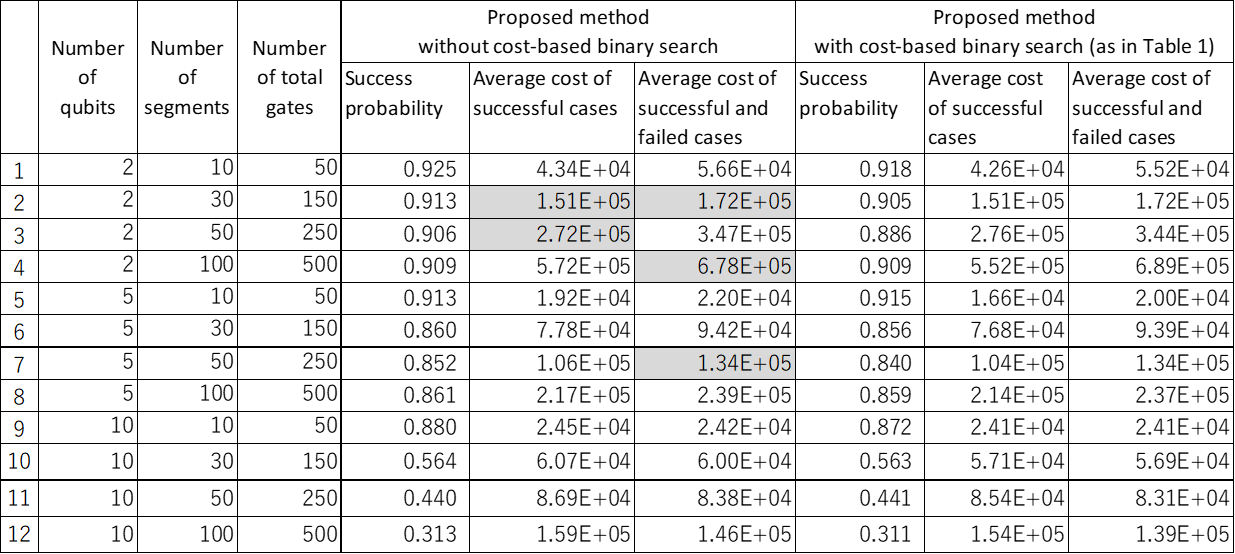}} %0 0 1786 205
\caption{Results with and without cost-based binary search}
\label{table02}
\end{table*}
Comparing the cases applying the cost-based binary tree shown in the right column of Table \ref{table02}, the average search cost increased except for the gray cells. These results confirm the effectiveness of cost-based binary search. However, the reduced cost by applying cost-based binary search is not so large. For the gray cells, the cost did not change or slightly increased. This suggests that the expected search cost of a search tree composed by naive binary search is not much larger than that of a cost-based binary search tree, and the cost of naive binary search may be lower than that of the cost-based binary search in some cases.
%Table \ref{table02}では、グレーのセルを除いて、cost-based binary searchを適用すると、平均探索コストが減少している。この結果から、cost-based binary searchの有効性を確認できる。ただし、cost-based binary searchの適用によって減少するコストはそれほど大きくないこと、およびグレーのセルのようにcost-based binary searchを適用してもコストが変わらない、または微増する場合がある。このことから、cost-based binary searchで構成したsearch treeのexpected search costは、naive binary search treeのそれよりも小さいが、その差は大きくないと考えられる。

\subsubsection{Effectiveness of early determination}
Table \ref{table03} lists the results when early determination was not applied. Since early determination relaxes the thresholds of the chi-square test, the search can proceed with less cost. We confirmed that the average search costs decreased by early determination in all cases, which means that early determination is very effective. We also confirmed that early determination contributes to improve the success probabilities in practice.
%Early determinationは、カイ二乗検定における判定の閾値を緩和することで、より少ないコストで探索を行えるようにする。Table \ref{table03}の全ての行について、early determinationを適用することで、平均探索コストが減少することを確認できた。つまりearly determinationは、探索コストの削減に効果的であることが分かる。また、上述の考察のとおり、early determinationの適用によってバグ位置特定の成功確率が向上することも確認できる。
\begin{table*}
\centering
\scalebox{0.6}{\includegraphics[bb=-100 0 889 250]{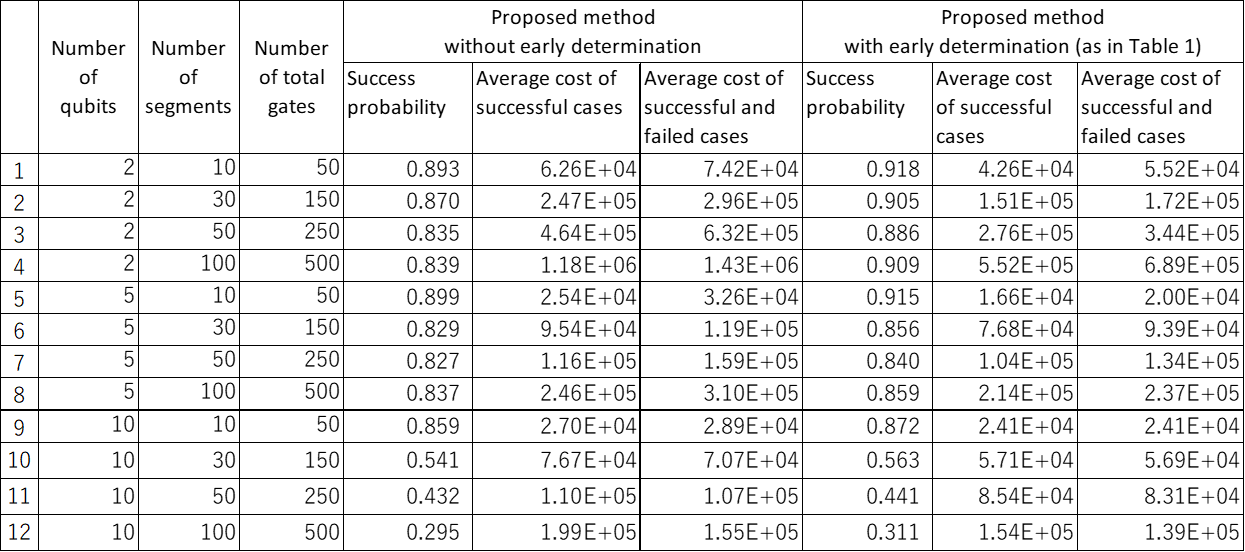}} %0 0 1786 205
\caption{Results with and without early determination}
\label{table03}
\end{table*}

\subsubsection{Effectiveness of finalization}
Table \ref{table04} lists the results when finalization was not applied. Since finalization adds measurements at the end of the search to prevent incorrect bug location, it increases the search cost. Instead, the success probability is expected to improve. In fact, from Table \ref{table04}, we confirmed that finalization improved the success probabilities except for those in the gray cells. In the gray cells, the success probabilities slightly decreased by applying finalization.
%さらに、finalizationを適用しなかった場合の実験結果をTable \ref{table04}に示す。Finalizationは、incorrectなバグ位置の特定を防ぐために、測定を追加し、十分な精度でテストを行うものである。そのため、Finalizationを適用することで探索コストは増加する。その代わりに、バグ位置特定の成功確率は向上することが見込まれる。実際にTable \ref{table04}、グレーのセルを除いて、Finalizationの適用によって成功確率が向上していることから、Finalizationの有効性を確認できる。一方、グレーのセルでは、Finalizationの適用によって成功確率が微減している。特に量子ビット数が10のケースでは、Finalizationによって測定を追加しても、量子ビット数に対して測定回数が不十分であり、判定誤りを是正できなかったものと考えられる。
\begin{table*}
\centering
\scalebox{0.6}{\includegraphics[bb=-100 0 889 250]{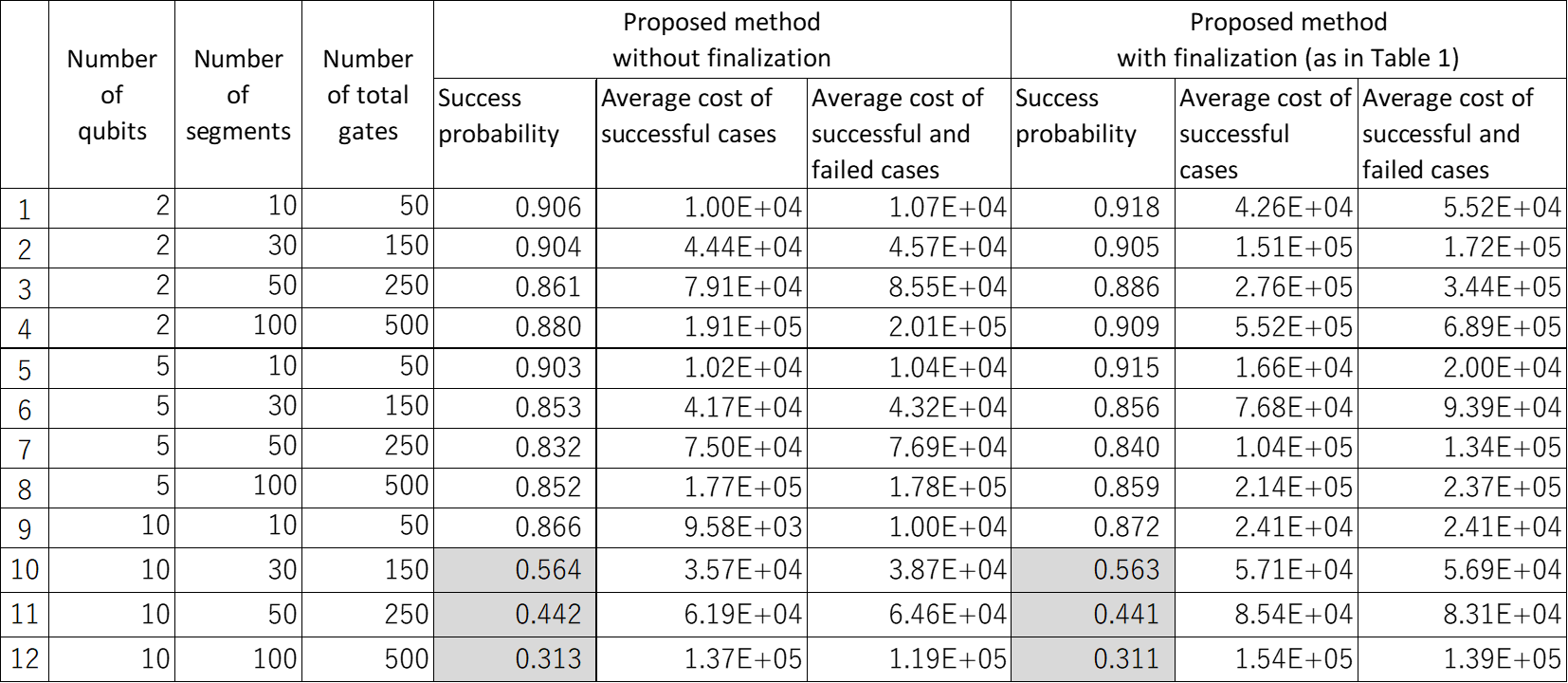}} %0 0 1786 205
\caption{Results with and without finalization}
\label{table04}
\end{table*}
%This shows that finalization is not effective in the case of 10 qubits. finalizationによって成功確率が低下するということは、early determinationによって正しい判定を下せていたが、finalizationによって測定を追加した結果、誤った判定に転じてしまったことを示唆している。提案手法では、RightEarlyまたはLeftEarlyと判定されるまで徐々に測定を追加する。この時、サンプリングの偏りによって誤ってRightEarlyまたはLeftEarlyと判定される可能性がある。そこでbuggy segmentを特定した後に、FinalizationによってそれぞれRightFinalizedまたはLeftFinalizedと判定されるまでさらに測定を追加する。測定を追加することで、サンプリングの偏りが平準化され、正しい判定に訂正されることが期待される。
This means that finalization is not effective for ten qubits. It is considered that a correct determination turned into a wrong determination as a result of additional measurements by finalization. With the proposed method, measurements are added until the determination of `RightEarly' or `LeftEarly' is obtained. At that time, it is possible that the sampling bias could make incorrect determination of `RightEarly' or `LeftEarly.' Therefore, further measurements are added by finalization until it is determined to be `RightFinalized' or `LeftFinalized.' By adding measurements, it is expected that the sampling bias will be leveled out and the incorrect determination updated to the correct one.

%量子ビット数$n=2 or 5$の場合、即ちカテゴリ数が少ない場合は、比較的少ないサンプリング数でもその偏りが平準化されやすいため、Finalizationの効果が期待できる。実際に、Table \ref{table04}においても、量子ビット数$n=2 or 5$ではFinalizationの適用によって成功確率が向上している。
When the number of qubits, i.e., the number of categories is not large, finalization is expected to be effective because the bias is likely to be leveled even with a relatively small number of measurements. In fact, we could confirm that finalization improved the success probabilities when the number of qubits was $n=2$ or $5$, as shown in Table \ref{table04}.
%一方でカテゴリ数が多い場合は（量子ビット数n=10）、偏りが平準化されるまで多くのサンプリング数が必要になる。ここでカイ二乗検定の特性に着目する。カイ二乗検定では、サンプル数の増加に伴ってカイ二乗値が増加しやすくなり、その結果、p値は小さくなりやすくなる。また、サンプル数が増加すると、検定力も大きくなりやすい\cite{helie2007understanding}。この特性は、バグがあるにもかかわらずRightEarly（恐らくバグがない）と誤って判定した際には有利に働く。Finalizationによってサンプル数を増加する過程で、p値は小さく、powerは大きくなる傾向にあるため、サンプリングの偏りが解消していなくても、バグがあるという正しい判定結果に更新される可能性が高い。
If the number of qubits is large, a large number of measurements will be required until the sampling bias is leveled out. We focus on the characteristics of the chi-square test. As the number of measurements increases, the chi-square value tends to increase; thus, the p-value tends to decrease. As the number of measurements increases, the power tends to increase \cite{helie2007understanding}. These characteristics are advantageous when we incorrectly obtain `RightEarly' (which means that there may be no bug) even though there is a bug. To obtain the correct determination, the p-value should be smaller and the power should be larger. This can be achieved by adding many measurements by finalization even if the sampling bias is not sufficiently leveled by the additional measurements. Therefore, finalization is advantageous to correct the incorrect determination of `RightEarly.'
%一方で、バグがないにもかかわらずLeftEarly（恐らくバグがある）と誤って判定した場合には、不利に働く。早期にサンプリングの偏りが解消せず、サンプル数が増加してしまうと、p値は減少傾向となり、閾値（T\_upper\_p）を超える可能性は低くなっていく。つまり、カテゴリ数が多い場合は、Finalizationによる訂正を期待できないケースがあるということである。さらに、バグがない時にRightEarlyと正しく判定したケースを考慮する。この場合、順当にいけばFinalizationによってLeftFinalizedに更新されるはずだが、Finalizationによって追加されたサンプリングの結果に偏りが生じ、Undeterminedなどに更新される場合もある。その場合、その偏りを解消するためにさらにサンプリングが必要になる。そしてサンプル数の増加に伴って、p値が閾値を超える可能性が低くなっていく。このように、カテゴリ数が多い場合は、LeftEarlyと正しく判定しているにもかかわらず、Finalizationによって誤判定に転落してしまうことがある。Finalizationを適用しない場合は、LeftEarlyによって判定をfinalizeするため、buggy segmentの特定に成功する。
When we incorrectly determine `LeftEarly' (there may be a bug) even though there is no bug, the characteristics of the chi-square test are disadvantageous. Assume that the sampling bias is not sufficiently leveled by the measurements added by finalization. %When the sampling bias is not sufficiently leveled due to the large number of categories, many measurements are added by finalization. 
At that time, the p-value tends to decrease from the characteristics described above. To update determination `LeftEarly' to `RightEarly' or `RightFinalized', the p-value is expected to increase. Therefore, finalization is disadvantageous to correct determination `LeftEarly.'
Consider the case in which we correctly determine `LeftEarly' when there is a bug. Finalization is expected to update `LeftEarly' to `LeftFinalized.' However, if the measurements added by finalization is biased, determination `LeftEarly' may be updated to something other than `LeftFinalized', such as `Undetermined.' At that time, more measurements will be added. If those additional measurements do not level the bias, the p-value becomes smaller, and it may even become impossible to return to `LeftEarly.' Thus, when the number of qubits is large, finalization may make an incorrect determination even though it was correctly determined as `LeftEarly' before executing finalization. If finalization is not applied, the presence of a bug is finally correctly determined by `LeftEarly.'
%以上より、カテゴリ数が多い場合は、Finalizationによる誤りの訂正を期待できないばかりか、正しい判定を覆してしまう可能性がある。そのため、量子ビット数$n=10$の場合には、the success probabilities are slightly decreased by applying finalization.ただし、finalizationによって測定のバイアスを平準化できる場合は、これらの不利な点は現れない。測定のバイアスは、測定回数を増加することで平準化される可能性が向上する。よって、finalizationの適用は量子ビット数$n$だけでなく、$m_{unit}$ and $M_{max}$に応じて判断すべきである。

In summary, when the number of qubits $n$ is large, finalization may not correct incorrect determinations and even overturn correct determinations. This is the reason the success probabilities decreased when $n=10$ in the gray cells in Table \ref{table04}. However, if the sampling bias is sufficiently leveled by the measurements added by finalization, these disadvantages will not appear. The possibility of leveling out the bias increases with the number of measurements defined as $m_{unit}$ and $M_{max}$. Therefore, the application of finalization should be determined from $n$, $m_{unit}$, and $M_{max}$. In fact, when we increased $m_{unit}$ and $M_{max}$ from $100$ and $100,000$ to $10,000$ and $10,000,000$, respectively, we could confirm that the number of gray cells decreased, which means finalization becomes more effective. The results are listed in Table \ref{table04_m_increased}.
\begin{table*}
\centering
\scalebox{0.6}{\includegraphics[bb=-60 0 889 250]{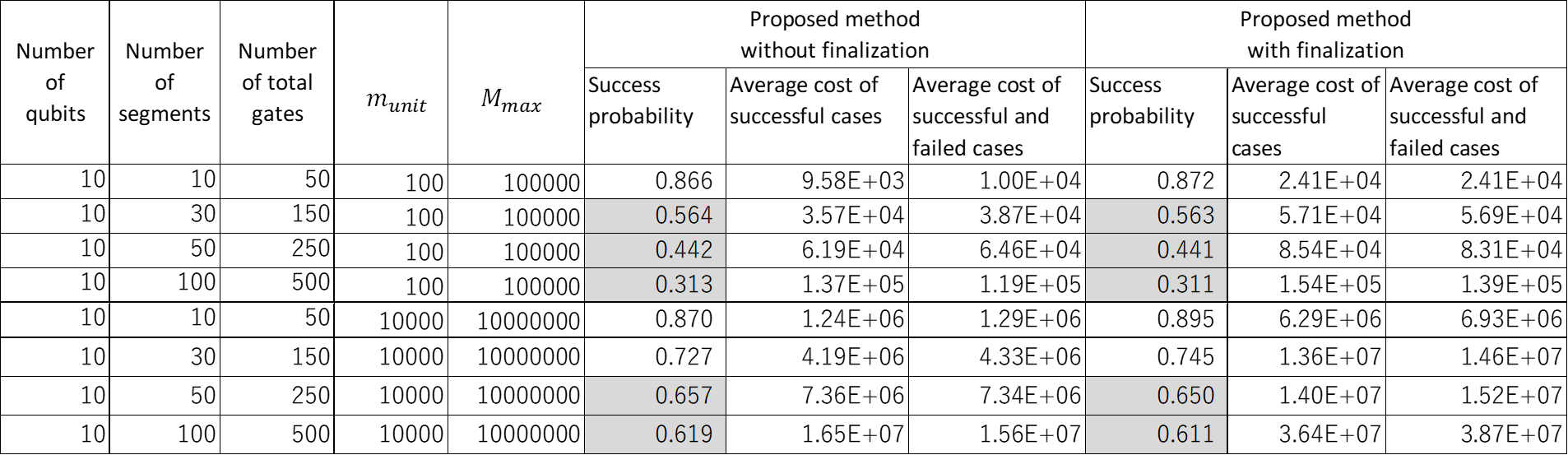}} %0 0 1786 205
\caption{Results with and without finalization before and after increasing measurements}
\label{table04_m_increased}
\end{table*}

\subsubsection{Effectiveness of looking back}
The results without looking back is shown in Table \ref{table05}. Looking back is expected to reduce search cost by detecting incorrect determinations made by early determination before finalization. Table \ref{table05} shows that the average search costs decrease by applying looking back except for the gray cells. If looking back suspects a correct determination, unnecessary measurements will be added and increase the search cost. This is the reason the average search costs increased in the gray cells.
%The gray cells indicate that there is a possibility that the correct determinations may be suspected by looking back. In that case, unnecessary measurements will be added and increase the search cost.
%最後に、looking backを適用しなかった場合の実験結果を、Table \ref{table05}に示す。Looking backは、early determinationによる判定の誤りを、finalizationにおいて検出するよりも先に検出することで、探索コストを削減するアプローチである。Table \ref{table05}ではグレーのセルを除いて、多くの場合、looking backを適用することで平均探索コストが減少することが確認できる。グレーのセルにおいてコストが増加しているのは、looking backによって正しい判定結果をsuspectしてしまい、本来不要な測定を行ったケースがあったためと考えられる。
\begin{table*}
\centering
\scalebox{0.6}{\includegraphics[bb=-100 0 889 250]{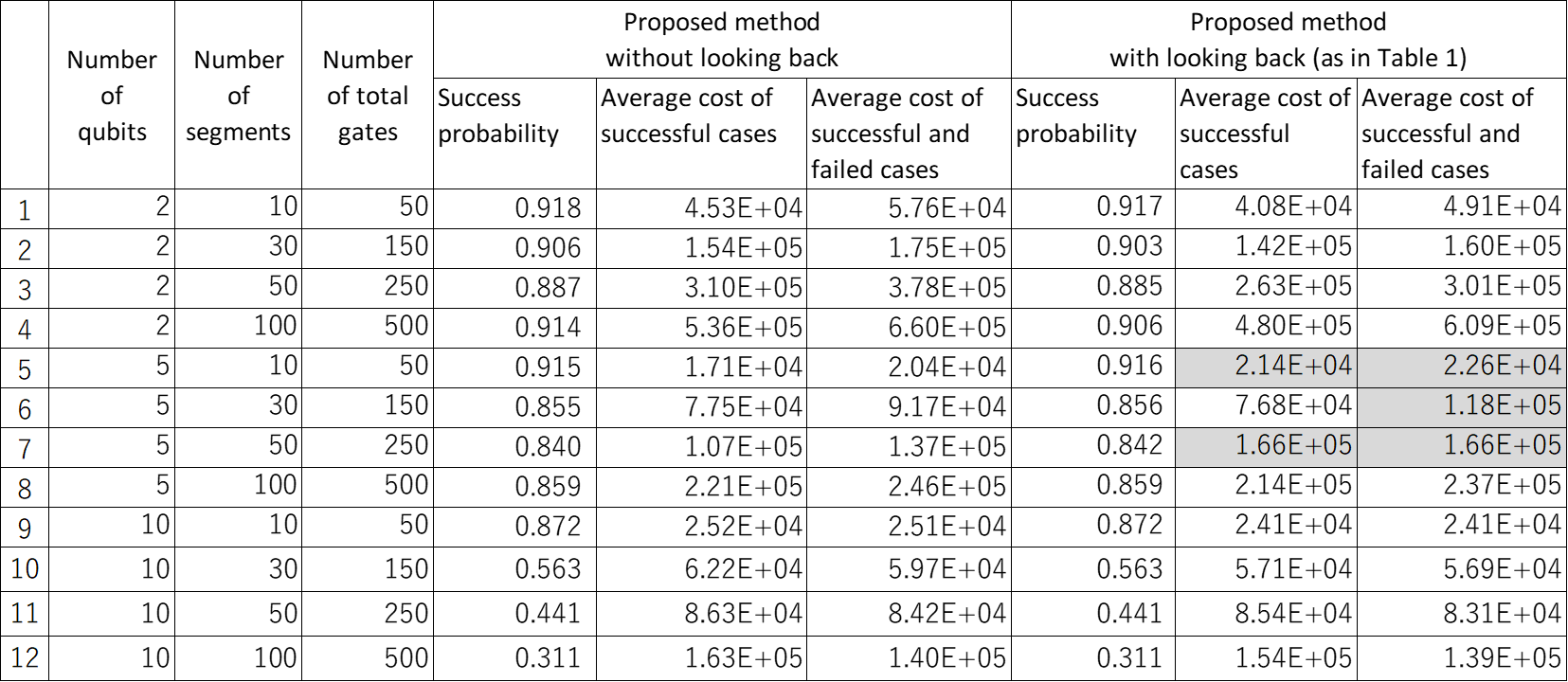}} %0 0 1786 205
\caption{Results with and without looking back}
\label{table05}
\end{table*}

%cost-based binary search, early determination, およびlooking backのいずれも探索コスト削減の効果があるが、各アプローチを適用したことで削減されるコストの大きさに着目すると、特にearly determinationが最も効果が高い。

\section{Discussion}\label{discuss}
\subsection{Theoretical Evaluation of Early Determination}\label{risk}
%ベイズ定理での分析
The basic idea of early determination is to take the risk of return instead of reducing the number of measurements. This section discusses the probability of return on the basis of Bayes' theorem. 
Assume that the quantum program is divided into $l$ segments and the search is executed from $n_1$ to $n_k$ with the path $p=[e_1^{d_1}, ..., e_i^{d_i}, ..., e_{k-1}^{d_{k-1}}]$. At node $n_i$, the segment $s_x$ ($1 \leq x \leq l-1$) is tested, and the executed sequence is denoted as $S_i$. When early determination is applied at each node, the Type I and Type I\hspace{-1.2pt}I error rates of a statistical test are denoted as $ \alpha $ and $ \beta $, respectively. Table \ref{table_hypothesis} shows the definitions of $ \alpha $ and $ \beta $.
\begin{table}
\centering
\scalebox{0.75}{\includegraphics[bb=0 0 619 180]{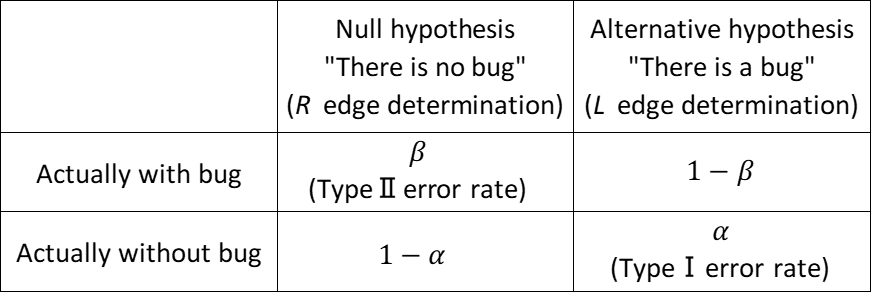}}
\caption{Definitions of Type I and Type I\hspace{-1.2pt}I error rates}
\label{table_hypothesis}
\end{table}

In the following, $e_i^{d_i} = e_i^{L}$ is discussed, but the same is applied for $e_i^{d_i} = e_i^{R}$. Let $w$ ($w \geq 1$) be the number of $L$ edges from $n_i$ to $n_k$. 
First, we consider the prior probability $P(B)$, which represents the probability that executed sequence $S_i$ does not have a bug. As in Section \ref{exp}, we assume the entire program contains only one bug and each segment has an equal chance of containing it. Let the length of $S_i$ be $x$. The probability that $S_i$ have a bug is $x/l$. Therefore, the $P(B)$ is expressed as $(l-x)/l$. 
Next, we consider the conditional probability $P(A|B)$ that the search follows path $p$ from $n_i$ to $n_k$ when $S_i$ contains no bug. Since we assume $e_i^{d_i} = e_i^{L}$, i.e., the search incorrectly determines that $S_i$ has a bug at node $n_i$, the executed sequences $S_{i+1}$ of $n_{i+1}$ to $S_{k-1}$ of $n_{k-1}$ are sub-sequences of $S_i$ and do not contain a bug. The probability of $L$ edge determination at a node from $n_i$ to $n_k$, that is, the probability of the incorrect determination of the presence of a bug is $ \alpha $. 
%from $e_i^L$ to $e_{k-1}^{d_{k-1}}$.
%Since we assume that the number of $L$ edges from $n_i$ to $n_k$, that is, determinations that there is a bug, is $w$, 
Since $w$ is the number of $L$ edges from $n_i$ to $n_k$, the joint probability of the incorrect $L$ edge determinations is $(\alpha)^w$. Similarly, the probability of the correct $R$ edge determination at a node from $n_i$ to $n_k$ is $(1 - \alpha)$. Their joint probability is $(1 - \alpha)^{(k-i)-w}$. Therefore, the conditional probability $P(A|B)$ that the search follows path $p$ from $n_i$ to $n_k$ when $S_i$ contains no bug is expressed as $(\alpha)^w (1 - \alpha)^{(k-i)-w}$.
The marginal probability $P(A)$ that the search reaches $n_k$ from $n_i$ along $p$ is then calculated as a sum of the probabilities for each bug location. For example, if the buggy segment is correctly narrowed down at node $n_k$, the probability that the search reaches $n_k$ from $n_i$ is $(1 - \beta)^w (1 - \alpha)^{(k-i)-w}$, where $(1 - \beta)^w$ and $(1 - \alpha)^{(k-i)-w}$ represent the joint probabilities of the correct $L$ edge determinations and the correct $R$ edge determinations, respectively. Otherwise, if there is one incorrect $R$ edge determination from $n_i$ to $n_k$, the probability is $ \beta (1 - \beta)^{w-1}(1 - \alpha)^{(k-i)-w}$. The probability that the bug is contained in each segment is $1/l$. Therefore, $P(A)$ is expressed as $((1 - \beta)^w (1 - \alpha)^{(k-i)-w})(1/l) + (\beta(1 - \beta)^{w-1}(1 - \alpha)^{(k-i)-w})(1/l) + ...$. For sake of simplicity, instead of the actual $P(A)$, we use the largest term $((1 - \beta)^w (1 - \alpha)^{(k-i)-w})(1/l)$ and omit the other terms.
%which is the probability that the buggy segment is correctly narrowed down at $n_{k}$. 
Finally, the posterior probability $P(B|A)$ that the search reaches $n_k$ from $n_i$ along $p$ when $S_i$ has no bug, that is, the probability that the search returns from $n_k$ to $n_i$ is expressed as
\begin{eqnarray*}
P(B|A) \simeq \frac{(\alpha)^w  (1 - \alpha)^{(k-i)-w}  ((l-x)/l)}{((1 - \beta)^w (1 - \alpha)^{(k-i)-w})(1/l)}.
\end{eqnarray*}
Since $ \alpha $ is much smaller than $1$, focusing on $(\alpha)^w$, we see that $P(B|A)$ decreases exponentially as $w$ increases. This indicates that the presence of $L$ edges after $e_i^L$ decreases the probability of returning to $n_i$. In Section \ref{early}, we interpreted this as the reinforcement relation of edges, which is the basis for early determination. If $w=1$, that is, only $R$ edges appear after $e_i^L$, the probability to return to $n_i$ does not decrease. In this case, the correctness of $e_i^L$ will be checked by additional tests, which is implemented as looking back mentioned in Section \ref{suspicious}.

\subsection{Limitations and Usefulness}
\subsubsection{Device implementation}
Various device implementations of quantum computers have been developed, e.g., superconducting \cite{sabre}, trapped-ion \cite{EfficientQubitRouting}, trapped-atom \cite{qubitmappingatom2022}, and silicon \cite{Lee_2020}\cite{Lee_2022}. Since our method is independent of hardware implementation, it can be applied to quantum programs running on any quantum computer.

\subsubsection{Testing method}
%[Z測定の制限]
With our method, an SBD method is internally used for quantum program testing. A statistical test has a certain probability of making an incorrect determination. Therefore, it is not always successful in locating a buggy segment. We used a simple testing method for this study, which is, measuring quantum states in the Z-basis and comparing the results with its oracle through the chi-square test. If we have bugs that do not appear as a difference in the amplitude of the Z-basis, such as a difference in phase, they cannot be detected with this testing method. To detect such bugs, for example, segments should be tested on the basis of measurement results in multiple bases. Quantum state tomography is also one of more rigorous methods, though it requires many measurements. Our method involves testing order of segments and reducing/adding the number of measurements, which does not limit the statistical testing method used internally. Therefore, if another statistical testing method is used to detect a bug, our method is still applicable.

%膨大な測定回数は現実的なのか
%実験の結果から、バギーセグメントの特定には、$10^6$程度の量子ゲートの実行を要することが分かった。量子ゲートの実行時間はナノ秒オーダーであり、$10^6$の量子ゲートを実行しても、xx秒で終了できる。実際には、初期化などにも時間がかかるため、これよりも長い時間がかかると想定されるが、提案手法によって、実用時間内でバギーセグメントを特定できると考える。
%In the experiments in Section \ref{exp}, we defined the testing cost as the number of quantum gates to be executed. 
Our experimental results indicate that it takes about $10^6$ quantum gate executions to locate a buggy segment. The execution time of a quantum gate is on the order of nanoseconds \cite{sung2021realization} \cite{kandala2021demonstration}, so $10^6$ quantum gates can be executed within one second. In fact, it will take longer because of the time required for initializations and measurements, but we believe that the proposed method can be used in practical time.

\subsubsection{Testing cost}
%[ノイズを考慮したテスティングコストの設計]
In the experiments in Section \ref{exp}, we defined the testing cost as the number of quantum gates to be executed. 
Quantum computers, which are expected to be commercialized in the near future, are called noisy intermediate-scale quantum (NISQ) computers and are affected by noise. When a quantum program is run on a NISQ computer, the measurement results are affected by noise and result in a larger statistical variance. Therefore, to ensure the same testing accuracy, more measurements are required than without noise. The test cost is even greater in a backward segment since the noise effect is accumulated from the gates before it, that is, the executed sequence. Therefore, when targeting NISQ computers, it is necessary to design the testing cost by taking into account the effect of noise. As a study focusing on the effects of noise in testing, Muqeet et al. proposed using machine learning to filter out the effects of noise and improve testing accuracy \cite{muqeet2023noise} \cite{muqeet2024mitigating}. This approach could be incorporated into our proposed method and reduce the cost of locating buggy segments in NISQ computers.

%[ゲートの種類を考慮したテスティングコストの設計]
The testing cost was defined without distinguishing single-qubit gates and two-qubit gates. However, since the implementations of single-qubit and two-qubit gates are essentially different, the testing cost should be designed considering the difference. In fault-tolerant quantum computing (FTQC), the next era of NISQ, errors caused by noise can be corrected. A logical quantum gate is achieved by configuring error-correcting code with multiple physical quantum gates \cite{PhysRevA.86.032324} \cite{google2023suppressing} \cite{bluvstein2024logical}. Since the number of physical quantum gates depends on the type of a logical quantum gate, it is necessary to design the testing cost considering the type of logical quantum gates consisting of each segment in FTQC.

\subsubsection{Test oracle}
%[Oracleを定義できない場合がある/解こうとしている問題の答えを知らない場合shapeしか分からない]
We assumed that the correct output quantum state of each segment could be expected as a test oracle. However, the developer often does not know the output state of the quantum program, in other words, the solution to a problem that the quantum program solves. In that case, it may be difficult to correctly expect the output state of a segment. For example, in the quantum program of Grover's algorithm shown in Figure \ref{segmentation}, the output state of segment $s_{15}$ is expected to be a unimodal categorical distribution where the solution of this quantum program (basis $|111>$) has the highest probability, about 80\%.
Thus, if the developer does not know the solution, a test oracle of segment $s_{15}$ cannot be given. %Therefore, we should assume that the solution is not known when testing the quantum program and it is difficult to give the oracle for $s_{15}$. 
This may affect the usefulness of the proposed method because it is more natural to assume that the solution is not known. However, even if the exact distribution cannot be given, we could often expect the shape of the distribution. In the above example, the distribution of $s_{15}$ can be expected as unimodal for any one of the categories. Although the testing accuracy will be reduced, we can test the segment.
%量子プログラムによって解こうとしている問題の答えが分からない場合、developerはセグメントの正しい出力状態をgiveできない可能性がある。例えば、Figure \ref{segmentation}に示したGroverのアルゴリズムの量子プログラムでは、segment $s_{15}$のoutput stateは、solutionである111が約80パーセントの確率でobtainされるunimodalなcategorical distributionであることが期待される。しかし、もしsolutionが111であると分かっているのであればそもそもこの量子プログラムは不要である。つまり、この量子プログラムのテストにおいては、solutionは分からないと仮定すべきである。よって、111が単峰となる分布をoracleとして与えるのは困難である。ただし、この場合は、「いずれかの値が単峰となる分布」をoracleとして与えることは可能である。その場合、精度は低下するがセグメントをテストできるため、その結果を用いて、提案手法によってバグ位置を特定することは可能である。

%[Oracleを定義できない場合がある/中途半端なところでセグメントを切るとオラクルを定義するのが難しい]
If the quantum program is divided into segments semantically, the oracle of a segment is easier to give than otherwise. For example, the quantum program in Figure \ref{segmentation} can be semantically split into parts of initialization, oracle, and diffusion. To prevent confusion, an oracle for testing and an oracle in Grover's algorithm are denoted as a test oracle and Grover's oracle, respectively. Segments from $s_2$ to $s_7$ correspond to the (first) Grover's oracle part. Since Grover's oracle only inverts the phase of the basis that indicates the solution, the absolute squares of the amplitudes are not changed. Therefore, if we know this specification of Grover's oracle, we can define the test oracle for segment $s_7$. However, the output state of $s_6$ is an intermediate state in the middle of Grover's oracle calculation. Even if we know the specification of Grover's oracle, it may be difficult to give the expected output state of $s_6$. 
%This means that the quantum program should be divided into segments so that the developer can define the test oracle for each segment.
%Figure \ref{segmentation}の量子回路は、意味的には、initialization、oracle、diffusionのパートに分けられる。To prevent confusion, oracles for testing and oracles in Grover's algorithm are denoted as test oracles and Grover's oracles, respectively. 例えば、segment $s_1$はinitializationのパートであり、$s_2$から$s_7$までが(1回目の)Grover's oracleのパートである。Grover's oracleはsolutionの基底の位相を反転させるだけであるため、test oracleとして与えられるthe absolute squares of the amplitudes are not changed. よって、Grover's oracleの仕様を知っていれば、segment $s_7$のtest oracleを定義できる。一方で、$s_2$から$s_6$のoutput statesは、Grover's oracleの途中の計算結果に相当する。そして、仕様を知っているだけでは、これらのtest oraclesを定義できない場合がある。よって、量子プログラムは任意にsegmentに分割できる訳ではなく、その分割方法はdeveloperのexpertiseに依存する。

%%%%%%%%%%%%%%%%%%%%%%%%%%%%%%%%%%%%%%%%%%%%%%%%%%%%%%%%%%%%%%%%%%%%%%%%%%%%%
%[Oracleを定義できない場合がある/nが大きくなるとオラクルを書き下せない]
Since a quantum state consisting of $n$ qubits is represented by $2^n$ basis states, as the number of qubits increases, it becomes difficult to describe the expected measurement results as a test oracle entirely. Therefore, the test oracles can be described only for a few bases, and the chi-square test is conducted using them. We present additional experimental results in Table \ref{table10} assuming that the number of bases that can be described is ten.
\begin{table*}
\centering
\scalebox{0.6}{\includegraphics[bb=-10 0 1591 300]{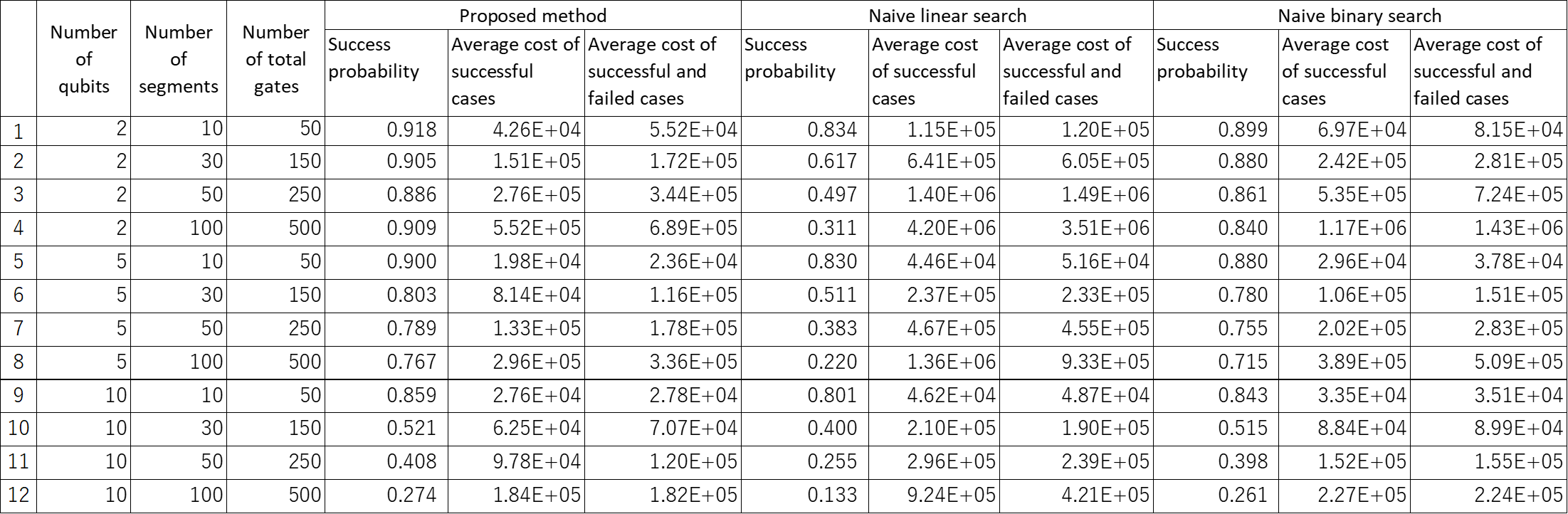}}
\caption{Results when 10 bases are described as test oracle}
\label{table10}
\end{table*}
For the qubit numbers $n=2$, the number of bases is $2^2=4$. Since it is smaller than 10, limiting the number of bases to 10 does not affect the results. When $n=5$ or $n=10$, the number of bases is $2^5=32$ or $2^{10}=1024$, respectively. From these bases, ten bases that could be measured, i.e., the absolute square of the amplitude is not zero, were randomly selected and used for the chi-square test.
%測定される可能性のある、即ち振幅の絶対値が0よりも大きい基底を10個選択し、それらをカイ二乗検定に使用した。
Table \ref{table10} shows that the average costs of the proposed method are lower than those of the comparison methods, which indicates that the proposed method is still more efficient than the comparison methods even if the number of bases that can be described as a test oracle is limited. 
If unselected bases (other than the ten selected bases) are observed by measurement, those measurement results are discarded and not used for the chi-squared test. If those unselected bases have a large difference in the absolute square of the amplitude between before and after the bug insertion, it will be difficult to detect bugs with the test. For this reason, the success probabilities in Table \ref{table10} tend to be lower than those in Table \ref{table01}.

\subsubsection{Thresholds in early determination}
Additional experimental results for more relaxed and stricter thresholds in early determination are listed in Table \ref{table06}. The results with more relaxed thresholds are shown on the left of the table, in which the thresholds of the p-value when determining the absence and presence of a bug are 0.5 and 0.2, respectively. The results on the right are with stricter thresholds that are 0.7 and 0.075, respectively. The results in the center are the same as those in Table \ref{table01}, in which the thresholds are 0.6 and 0.1, respectively.
\begin{table*}
\centering
\scalebox{0.6}{\includegraphics[bb=-10 0 1591 300]{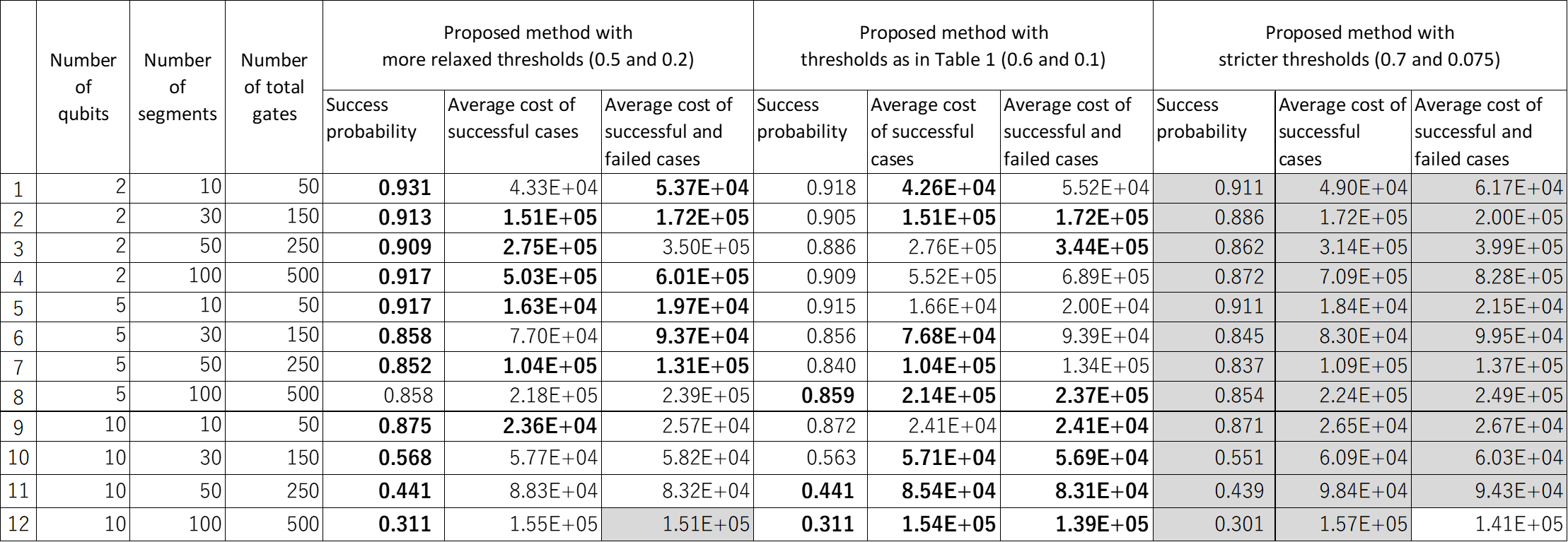}}
\caption{Results when changing thresholds of early determination}
\label{table06}
\end{table*}
%Table \ref{table06}では、3通りの閾値のうち、成功確率が最も高いものを太字で記述し、最も低いものをグレーで色付けしている。同様に、平均コストが最も低いものを太字で記述し、最も高いものをグレーで色付けしている。つまり、太字は良い結果、グレーの色付けは悪い結果を意味する。Table \ref{table06}より、小規模な量子プログラムの場合は、閾値を緩和した（0.5 and 0.2）方が良い結果（high success probability and less cost）が得られやすく、回路規模が増大するにつれて、徐々にstricterな閾値の方が良くなっていく傾向が見て取れる。この結果から、提案手法をより有効に使用するためには、回路の規模に応じて閾値を適切に変更する必要があることが分かる。
In Table \ref{table06}, the highest success probabilities are in bold and the lowest are in gray. Similarly, the lowest average costs are in bold and the the highest are in gray. Table \ref{table06} shows that for small quantum programs, more relaxed thresholds (0.5 and 0.2) tend to give better results (high success probability and less cost). As the circuit size increases, the results become better with normal thresholds. In the cases of the stricter threshold, both the success probabilities and the average costs are the worst in almost all cases. This indicates that early determination is not effective if the thresholds are too strict. However, for larger quantum circuits than those used in this experiment, the stricter thresholds may lead to better results. Thus, it is necessary to change the values of the thresholds appropriately in accordance with the circuit size.

%%%%%%%%%%%%%%%%%%%%%%%%%%%%%%%%%%%%%%%%%%%%%%%%%%%%%%%%%%%%%%%%%%%%%%%%%%%%%
%TODO
\subsubsection{Multiple buggy segments}
In the experiment in Section \ref{exp}, one buggy segment was made by replacing a quantum gate of the segment. The proposed method is useful when multiple bugs are included in a quantum program. For example, if there are two buggy segments, the proposed method will locate the forward buggy segment first. By applying the proposed method again after fixing the located bug, the backward buggy segment will be located. Similarly, if multiple bugs are included, the proposed method can be applied iteratively until locating and removing all bugs. The results of applying the proposed method with this iterative procedure to quantum programs containing two buggy segments are listed in Table \ref{table07}. 
%バギーセグメントが1つの場合は、それを特定できたかどうかの成功確率を評価した。しかし、バギーセグメントが2つの場合は、片方のバギーセグメントの特定に成功し、もう片方の特定に失敗する場合もある。そこで我々は、ある量子回路において2つのバギーセグメントを両方特定できた確率、およびそれぞれのバギーセグメントを正しく特定できた確率、即ちRecallを計算した。
%Since a quantum program contains two buggy segments, there are three cases: when both buggy segments are successfully located, when only one buggy segment is located, and when none of the buggy segments are located. 
We calculated the probabilities of successfully locating both buggy segments, the average costs for the successful cases, and the average costs for both cases of success and failure.
% and the probabilities of locating each buggy segment, i.e., recall.
\begin{table*}
\centering
\scalebox{0.55}{\includegraphics[bb=0 0 1591 300]{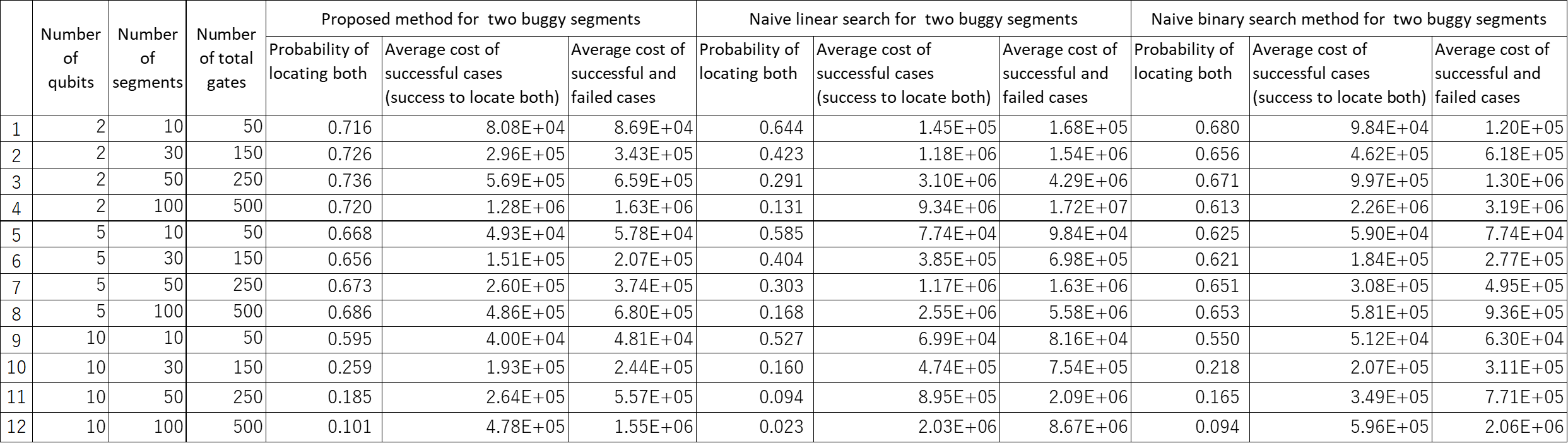}}
\caption{Results when applying to quantum programs with two buggy segments}
\label{table07}
\end{table*}
%2つのバグを含むプログラムにも有効であることが確認できた。2つのbuggy segmentを両方とも正しく特定できる確率は、buggy segmentが一つの場合の合成確率となるため、Table \ref{table01}と比較して、成功確率は低くなっている。
From these results, we could confirm that the proposed method can also be applied for quantum programs containing multiple bugs. The probabilities of locating both buggy segments were lower than the success probabilities for single buggy segments, as shown in Table \ref{table01}. That is because they are calculated as composite probabilities of success probabilities with single buggy segments. 
%The probabilities of locating each buggy segment were higher than the success probabilities for single buggy segments. In the iterative procedure for multiple buggy segments, even if a non-buggy segment is wrongly identified as a buggy one, the procedure continues to search for buggy segments. For a single buggy segment, however, the procedure terminates when one segment is located. This means that there is a higher chance of locating the buggy segments by the procedure for multiple buggy segments than for a single buggy segment. Therefore, the probability of locating each buggy segment is higher than the success probabilities in Table \ref{table01}. 
Table \ref{table07} also shows that the average costs of the proposed method are lower than those of the comparison methods, that is, it is more efficient even for multiple buggy segments.

%left sequenceに一つでもバグがあるにもかかわらずそれを見逃すと、後方のセグメント列にはバグがないのに、バグを探すことになり、その結果、最後尾または最前列のセグメントにバグがあると誤判定する。→これはバグが1つの場合も同じなので、理由として記述しない

%%%%%%%%%%%%%%%%%%%%%%%%%%%%%%%%%%%%%%%%%%%%%%%%%%%%%%%%%%%%%%%%%%%%%%%%%%%%%
\subsubsection{Filtering of quantum programs}
%Section \ref{exp}で述べたとおり、実験用の量子プログラムを作成する際に、バグの挿入前後のthe absolute square of the amplitudeの差を計算し、その差が小さいプログラムは、フィルターアウトした。これにより、実験に使用する量子プログラムは、カイ二乗検定でバグを検知しやすい量子プログラムになるようにした。しかし実際の開発現場には、バグの検知が容易なプログラムのみが存在するわけではない。そこで、カイ二乗検定でバグを検知しにくい量子プログラムをフィルターせずに、1000の量子プログラムを生成し、同様の実験を行った。結果をTable \ref{table08}に示す。
As described in Section \ref{exp}, when creating the quantum programs for the experiments, quantum programs with a small difference in the absolute square of the amplitude between before and after the bug insertion were excluded. This ensured that a bug in the quantum programs was likely to be detected with the chi-square test. 
%This exclusion has no problem in our exepriments because we intended to evaluate the proposed method comparing to the naive methods.
%the differences in the absolute squares of the amplitudes before and after the insertion of the bug were calculated. If the differences were small, the quantum program was excluded. 
If it is difficult to detect the bug with the chi-square test, both the proposed method and naive methods are likely to fail in locating the buggy segment. Therefore, it is difficult to compare their efficiency. Therefore, the quantum programs were filtered in the experiments, as described in Section \ref{exp}. However, in practice, this filtering is not natural. If the quantum programs are not filtered, the success probabilities will be lower than in Table \ref{table01}. To confirm this, we newly generated quantum programs without filtering and conducted the same experiment. The results are listed in Table \ref{table08}.
\begin{table*}
\centering
\scalebox{0.6}{\includegraphics[bb=-30 0 1591 300]{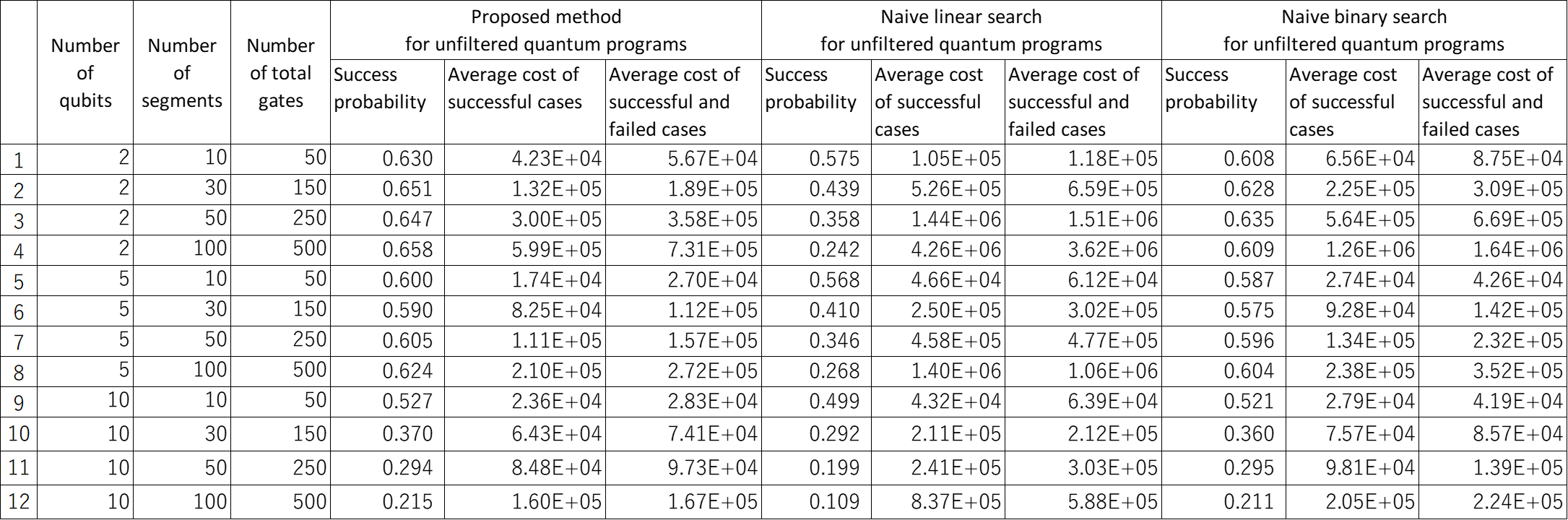}}
\caption{Results for quantum programs whose differences of amplitudes are small}
\label{table08}
\end{table*}
The search costs are almost the same as in Table \ref{table01}, but the success probabilities became lower than in Table \ref{table01} as expected. Table \ref{table08} also shows that even if the quantum programs are not filtered, the average costs of the proposed method are lower than those of the comparison methods. %However, the success probabilities were smaller than in Table \ref{table01}. 
This indicates that the proposed method is still efficient in actual development. %However, when we use it in industrial development, we should understand that the accuracy of locating buggy segments is not very high.

\begin{comment}
%%%%%%%%%%%%%%%%%%%%%%%%%%%%%%%%%%%%%%%%%%%%%%%%%%%%%%%%%%%%%%%%%%%%%%%%%%%%%
As described in Section \ref{sec-tree}, in our prior work \cite{sato2024locating}, the middle element $s_x$ was selected so that the highest total testing costs for searching the left and right sequences are as similar as possible. However, the average search cost was not minimized in that way. Therefore, we presented the updated algorithm in this paper. To evaluate the effectiveness of the current method, we compare the experimental results before and after updating the algorithm to select the middle element.
%その効果を評価するため、ミドルエレメントの選択方法をアップデートする前とした後の結果を比較する。
\begin{table*}
\centering
\scalebox{0.6}{\includegraphics[bb=-100 0 889 250]{figures/table09.png}}
\caption{Results before and after updating how to select the middle element}
\label{table09}
\end{table*}
Table \ref{table09} shows that the average costs are reduced by the update except for the gray cells. The method for constructing search trees proposed in this paper is better than our prior one in most cases.
\end{comment}

\section{Conclusion}\label{conclusion}
%We presented four characteristics that should be considered to locate buggy segments of quantum programs on a quantum computer. 
We presented an efficient bug-locating method consisting of cost-based binary search, early determination, finalization, and looking back. We also presented experimental results and theoretical evaluation to show the efficiency of the proposed method and discussed its usefulness and limitations. 
%[古典でシミュレーション実行できないサイズの量子プログラムで実証]

Future work will include the following studies. Our method targets large-scale quantum programs that cannot be simulated on a classical computer. Therefore, we should conduct experiments with larger quantum programs on an actual quantum computer. Our research group has been developing a silicon quantum computer \cite{Lee_2020}\cite{Lee_2022} \cite{utsugi2023single} \cite{mizuno2023quantum} \cite{sekiguchiSiQEW} \cite{sato2024generating}, but we do not yet have an environment to conduct those experiments. 
%We will also demonstrate the usefulness of the proposed method in the entire debugging process, e.g., whether the proposed method can locate multiple bugs. The proposed method should locate the most forward buggy segment. By applying the proposed method again after fixing the bug, another segment including a bug will be located.
%Improvements to the proposed method is also included for future plans. 

%[リスクに基づいて精度の閾値を動的に変更]
%Section \ref{risk}に示したように、カレントノードから、その上位の各ノードへの手戻りリスクを計算できる。上位ノードへの手戻りリスクが高まってきた場合の対策として、Looking Backを提案したが、更なる対策も考えられる。例えば、Lエッジの後に、N回のRエッジが続いており、カレントノードにてLエッジの判定が出たとする。その場合、上位のLエッジへの長さNの手戻りのリスクを大きく軽減できる。しかし、カレントノードの判定が間違っていた場合は、誤ってリスクを大きく削減することになってしまう。よって、上位ノードへの手戻りリスクを大きく軽減するような判定を行う場合は、indicatorの閾値を緩和せずに、十分な精度で判定を行う。これにより、手戻りのリスクを間違って軽減してしまい、誤りがあるのに、Looking Backが発動しないことを防止できる。
As discussed in Section \ref{risk}, the return risk from the current node to each upper node in the search tree can be calculated. Looking back is an approach for mitigating the return risk when the risk increases by the successive occurrence of the same type of edge ($R$ or $L$ edge). The risk can also be controlled by dynamically changing the thresholds of the accuracy indicators. For example, suppose that the threshold $D$ of looking back is defined as 5 and an $L$ edge at node $n_1$ is followed by three $R$ edges from nodes $n_2$ to $n_4$. If an $L$ edge is made at the current node $n_5$, the determination at $n_1$ will no longer be suspected by looking back due to the determination at $n_5$. Therefore, the determination at $n_5$ is important and should be made carefully. By dynamically adjusting the thresholds of the accuracy indicators, the test at $n_5$ can be conducted with sufficient accuracy, which prevents incorrect determinations.

%[Bug patterns based on subroutines]
Many quantum algorithms consist of a combination of specific subroutines and their repetition. In Figure \ref{segmentation}, for example, there are many similar segments in the quantum program. Assume that a quantum program is defined on the basis of subroutines. One of them has a bug and it is used multiple times in the entire quantum program. When a buggy segment in the subroutine is located, it indicates another buggy segment included in the same subroutine called elsewhere. By focusing on the structure of the quantum algorithm, the proposed method could be made more efficient. 

%[却下]The efficiency of the proposed method when used with other testing methods described in Section \ref{relwork} should also be evaluated.
%・今回の実験では、バグが各セグメントに存在する確率は同一としたが、
%　本来、バグが各セグメントに存在する確率は、セグメントのゲート数に比例する。
%　その場合、コスト期待値を計算する際に、その確率の違いを考慮する必要がある。

% use section* for acknowledgment
%\ifCLASSOPTIONcompsoc
  % The Computer Society usually uses the plural form
%  \section*{Acknowledgments}
%\else
  % regular IEEE prefers the singular form
%  \section*{Acknowledgment}
%\fi

%The authors would like to thank...

% Can use something like this to put references on a page
% by themselves when using endfloat and the captionsoff option.
\ifCLASSOPTIONcaptionsoff
  \newpage
\fi

% trigger a \newpage just before the given reference
% number - used to balance the columns on the last page
% adjust value as needed - may need to be readjusted if
% the document is modified later
%\IEEEtriggeratref{8}
% The "triggered" command can be changed if desired:
%\IEEEtriggercmd{\enlargethispage{-5in}}

% references section

% can use a bibliography generated by BibTeX as a .bbl file
% BibTeX documentation can be easily obtained at:
% http://mirror.ctan.org/biblio/bibtex/contrib/doc/
% The IEEEtran BibTeX style support page is at:
% http://www.michaelshell.org/tex/ieeetran/bibtex/
%\bibliographystyle{IEEEtran}
% argument is your BibTeX string definitions and bibliography database(s)
%\bibliography{IEEEabrv,../bib/paper}
%
% <OR> manually copy in the resultant .bbl file
% set second argument of \begin to the number of references
% (used to reserve space for the reference number labels box)

\bibliographystyle{IEEEtran}
\bibliography{debugnavi}

\end{document}